\documentclass[a4paper,amsmath,amssymb,jcp,aip,preprint,unsortedaaddress]{revtex4-1}

\pdfoutput=1
\usepackage{aicescover}
\usepackage{natbib}
\usepackage[T1]{fontenc}
\usepackage[latin1]{inputenc}
\usepackage{color,graphicx}
\usepackage{array}
\usepackage{tabularx}
\newcolumntype{L}[1]{>{\raggedright\arraybackslash}m{#1}} % linksbündig mit Breitenangabe
\newcolumntype{C}[1]{>{\centering\arraybackslash}m{#1}} % zentriert mit Breitenangabe
\newcolumntype{R}[1]{>{\raggedleft\arraybackslash}m{#1}} % rechtsbündig mit Breitenangabe
\usepackage{pgfplots}
\pgfplotsset{compat=1.5}
\usepackage{tikz}
\usepackage{setspace}
\usepackage{hyperref}

\everymath{\displaystyle}
\begin{document}
\pagenumbering{roman}
\aicescovertitle{Multilevel Summation for Dispersion: A Linear-Time Algorithm for $r^{-6}$ Potentials}
\aicescoverauthor{Daniel Tameling \and Paul Springer \and Paolo Bientinesi \and Ahmed E. Ismail}
\aicescoverpage

\pagenumbering{arabic}

% Metadata and abstract

\title{Multilevel Summation for Dispersion: A Linear-Time Algorithm for \texorpdfstring{$r^{-6}$}{r-6} Potentials}
\author{Daniel Tameling}
\email{tameling@aices.rwth-aachen.de}
\affiliation{AICES Graduate School, RWTH Aachen University, Schinkelstr.\ 2, 52062 Aachen, Germany}
\affiliation{Aachener Verfahrenstechnik: Molecular Simulations and Transformations, Faculty of Mechanical Engineering, RWTH Aachen University, Schinkelstr.\ 2, 52062 Aachen, Germany}
\author{Paul Springer}
\author{Paolo Bientinesi}
\affiliation{AICES Graduate School, RWTH Aachen University, Schinkelstr.\ 2, 52062 Aachen, Germany}
\author{Ahmed E. Ismail}
\email{aei@alum.mit.edu}
\affiliation{AICES Graduate School, RWTH Aachen University, Schinkelstr.\ 2, 52062 Aachen, Germany}
\affiliation{Aachener Verfahrenstechnik: Molecular Simulations and Transformations, Faculty of Mechanical Engineering, RWTH Aachen University, Schinkelstr.\ 2, 52062 Aachen, Germany}

\date{\today}

\begin{abstract}
We have extended the multilevel summation (MLS) method, originally developed to evaluate long-range Coulombic interactions in molecular dynamics (MD) simulations [Skeel et al., \emph{J.\ Comput.\ Chem.}, \textbf{23}, 673 (2002)], to handle dispersion interactions.
While dispersion potentials are formally short-ranged, accurate calculation of forces and energies in interfacial and inhomogeneous systems require long-range methods.
The MLS method offers some significant advantages compared to the particle-particle particle-mesh and smooth particle mesh Ewald methods.
Unlike mesh-based Ewald methods, MLS does not use fast Fourier transforms and is thus not limited by communication and bandwidth concerns.
In addition, it scales linearly in the number of particles, as compared with the $\mathcal{O}(N \log N)$ complexity of the mesh-based Ewald methods.
While the structure of the MLS method is invariant for different potentials, every algorithmic step had to be adapted to accommodate the $r^{-6}$ form of the dispersion interactions.
In addition, we have derived error bounds, similar to those obtained by Hardy for the electrostatic MLS [Hardy, Ph.D. thesis, University of Illinois at Urbana-Champaign (2006)].
Using a prototype implementation, we have demonstrated the linear scaling of the MLS method for dispersion, and present results establishing the accuracy and efficiency of the method.

\begin{spacing}{.8}
  \small{
    \vspace*{4mm}
    \noindent
    Copyright (2014) American Institute of Physics. This article may be downloaded for personal use only. Any other use requires prior permission of the author and the American Institute of Physics.\\
    The following article appeared in The Journal of Chemical Physics and may be found at \url{http://dx.doi.org/10.1063/1.4857735}.
  }
\end{spacing}
\end{abstract}

\date{09 January 2013}
\maketitle

% Introduction

\section{Introduction}

The most expensive step in almost all molecular dynamics (MD) simulations is calculating the forces acting on the particles being studied, as well as their energies.
Among the various short-ranged and long-ranged forces included, among the most frequently encountered are dispersion forces, weakly attractive interactions
originating from induced dipoles caused by instantaneous asymmetry in the
arrangement of electrons around the nuclei of atoms.
The only attractive interaction between all pairs of atoms, dispersion is typically included in pairwise interactions between non-bonded particles.
Dispersion potentials are normally treated as proportional to $r^{-6}$, where
$r$ is the distance between two atoms. Such forms are found in many commonly
used potentials, such as the Lennard-Jones (LJ), Buckingham (exp-6), and Born
models.\cite{Stone1997,Tosi1964,Fumi1964}
Dispersion interactions are
important for many physical effects and material properties,
such as surface tension, stiffness, and adhesion, but can also affect
fundamental properties such as density and pressure.~\cite{Israelachvili2011}
Accurate representation of dispersion in MD simulations is therefore often
of vital importance. In this paper, we show how the multilevel summation
(MLS) method of Hardy et al.,\cite{Skeel2002, Hardy2006, Hardy2009} a recently
developed linear-time, long-range solver for electrostatics, can be extended to
dispersion interactions.

Since $r^{-6}$ is mathematically short-ranged, dispersion interactions have
normally been truncated beyond a cutoff $r_c$.~\cite{Griebel2007}
Technically, this approach is accurate only for homogeneous systems,
in which cutoff effects are mitigated by the isotropic structure.
For inhomogeneous systems, including interfacial systems and
systems with sharp concentration gradients, the introduction of cutoffs
leads to significant inaccuracies in both the thermodynamic properties and
the dynamics;~\cite{Veld:2007ip, Isele-Holder2012}
these inaccuracies have been
demonstrated in numerous interfacial simulations.~\cite{Chapela1977,
  Guo1997, Wen2005, janecek2006, Ismail2006, Ismail2007}
While post-processing methods have been devised to correct some of these problems,~\cite{Chapela1977, Guo1997, janecek2006} such treatments do
not change the underlying dynamics, thereby leading to errors in quantities
such as densities and diffusion coefficients.
For high-accuracy simulations, an ``on-line'' long-range treatment
of dispersion interactions is therefore necessary.

For typical MD simulations, the direct evaluation of long-ranged non-bonded
pairwise interactions remains impractical because of its
$\mathcal{O}\left(N^2\right)$ complexity, where $N$ is the number of
particles; as a consequence, a variety of solvers were developed to accurately
approximate long-range interactions.
These long-range solvers can be classified into two different categories,
namely, \emph{tree methods} and \emph{grid-based methods}.

Tree-based methods include the Barnes-Hut technique~\cite{Barnes1986}
and the fast multipole method (FMM).~\cite{Greengard:1987:fa} The FMM
shares two appealing properties with the MLS method: the optimal
linear scaling in the number of particles, and a natural decomposition
of length scales for multiple-time-step methods.  However, the FMM
comes with disadvantages too: for instance, the discontinuities
introduced by multipole expansions make the potential non-smooth,
which can cause problems in MD simulations.\cite{Bishop1997} By
contrast, the potentials in the MLS are always smooth, and while the
FMM is more accurate at lower cost, its non-smooth potential makes it
less efficient than the MLS.\cite{Skeel2002} Moreover, the FMM becomes
more expensive when used with periodic boundaries, although efforts to
optimize multipole expansions using periodic boundaries are
ongoing.\cite{lorenzen2012} However, as most MD simulations
incorporate periodic boundary conditions, and as it is inherently
difficult to craft an efficient implementation, the FMM has yet to be
incorporated into any major open-source molecular dynamics packages.

For long-range calculations, modern MD codes primarily use grid-based methods;
in particular, most of them are based on the Ewald method~\cite{Ewald:1921},
which splits the sum of interactions into a local term treated in real space
and a long-range term treated in Fourier space.
Unlike the original sum, which is only conditionally convergent,
both of the new terms are absolutely convergent.
While a fully optimized Ewald method can scale as $\mathcal{O}\left(N^{3/2}\right)$;\cite{Perram1988}
it remains impractical for systems with more than a few thousand particles.
Mesh-based adaptations of the Ewald method, such as the particle-particle particle-mesh (PPPM)
method,\cite{Hockney:1988} the particle-mesh Ewald (PME)
method,\cite{Darden:1993in} and the smooth particle mesh Ewald (SPME)
method,\cite{Essmann:1995vj} further reduce the complexity to
$\mathcal{O}\left(N \log N\right)$,  partly by using
fast Fourier Transforms (FFT's).
However, for extremely large systems running on tens of thousands of processors,
the cost of the all-to-all communications can dominate the cost of the algorithm.\cite{Isele2013}

The MLS method belongs to the class of grid-based methods, and, in
contrast to the mesh-based Ewald methods, entirely avoids FFTs and
their inherent disadvantages, while offering greater flexibility in
handling boundary conditions: periodic, non-periodic and mixed
boundary conditions are all permitted, and have negligible effect on
performance. Initially introduced for the computation of integral
transforms,\cite{Brandt1990} and later used for charge-dipole
interactions in two dimensions,\cite{Sandak2001} the MLS
method revolves around hierarchical matrix expansions. Later
work extended its use to MD simulations.\cite{Skeel2002} Hardy
significantly strengthened both the theory and the
implementation,\cite{Hardy2006} deriving strict and computational
error bounds, and mitigating concerns about the accuracy of the MLS
method previously raised.~\cite{Skeel2002}

Given the historical tendency to handle dispersion potentials via truncation,
it is unsurprising that all MD long-range methods were originally developed
for electrostatic potentials, and only later applied to other interactions, such as dispersion.
Williams treated formally short-ranged potentials, and in particular
dispersion, with a long-range method.\cite{williams1971}
The first implementation of a long-range dispersion algorithm using a mesh-based solver appears to have been proposed by Shi and co-workers,\cite{Shi2006} who constructed a PPPM version.
More recently,
in 't Veld et al.\cite{Veld:2007ip} and Isele-Holder et al.\cite{Isele-Holder2012}
provided implementations of the classical Ewald and PPPM methods for
dispersion potentials in the open-source molecular dynamics package
LAMMPS,\cite{Plimpton1995}
and a particle-mesh Ewald version of the algorithm has recently been added to Gromacs as well.\cite{Wennberg:2013bx}

The work described in this article represents, to our knowledge, the first attempt at
implementing MLS for dispersion calculations. In~\autoref{section: derivation
  MSM}, we generalize the existing algorithm, providing a potential
independent formulation where possible, and the necessary details for
dispersion, highlighting the differences to the electrostatic case.
To demonstrate its correctness and functionality,
in~\autoref{section: results} we apply the algorithm to LJ fluids,
we present preliminary error bounds,
and discuss the performance and complexity of the method in detail,
including a confirmation of the linear scaling of the method for
dispersion interactions.
We briefly summarize our results in \autoref{section: conclusions}.

%%% Local Variables:
%%% mode: latex
%%% TeX-master: "DanielPaper01"
%%% End:

% Mathematical Formulation

\section{Mathematical Formulation of the Multilevel Summation Method}
\label{section: derivation MSM}

\subsection{The Multilevel Summation Algorithm for Dispersion}
\label{s:mlsalgorithm}

We begin by demonstrating how to formulate pair potentials, so that the multilevel summation can be used.
Thereafter, we present a short introduction to the multilevel summation algorithm.
In particular, we show the close correspondence between the matrix structure on a formal mathematical level with the grid structure used in the implementation of the algorithm.
We also show how to apply the algorithm to arbitrary functions of the form
$r^{-p}$, with special focus on dispersion potentials, where $p = 6$.
To make our presentation self-contained, where appropriate we have included generalized versions of the original derivations by Skeel, Hardy and co-workers.~\cite{Skeel2002,Hardy2006,Hardy2009}

\subsubsection{Rewriting pair potentials for Multilevel Summation}

Non-bonded potential energy contributions are typically written as a
sum over all two-particle interactions. The corresponding potential $V$
can in principle be written as
\begin{equation}
V = \boldsymbol{\xi}^T \mathbf{G} \boldsymbol{\xi},
\label{eq:pot-vec-mat-vec}
\end{equation}
where $\xi_iG_{ij}\xi_j$ represents the interaction between particles
$i$ and $j$.  For position-dependent calculations, the distances
$r_{ij}$ between particle $i$ and particle $j$ are normally included
in the matrix $\mathbf{G}$, while the components of $\boldsymbol{\xi}$
are usually intrinsic properties of the particle.  For instance, for
electrostatic interactions $G_{ij} = r_{ij}^{-1}$ and $\xi_i = q_i$, where $q_i$ is the
charge of atom $i$.\cite{Skeel2002,Hardy2006}
For dispersion potentials, the components of the matrix $\mathbf{G}$ can be defined as:
\begin{equation}
  G_{ij} =
  \begin{cases}
    r_{ij}^{-6}, & i \ne j \\
    0, & i = j
  \end{cases}.
\end{equation}
Following Refs.~\onlinecite{Skeel2002, Hardy2006}, we call the function $r^{-6}$ a \emph{kernel}.
The definition of $\boldsymbol{\xi}$ is somewhat more complicated, as it depends on the choice of mixing rules.
As an example, for the LJ potential, we can write $\boldsymbol{\xi_i} = \mathbf{f}(\varepsilon_i, \sigma_i)$, where $\varepsilon_i$ and $\sigma_i$ are the well depth and zero, respectively.
Note that this definition implies the individual rows of $\boldsymbol{\xi}$ may themselves be vectors, so that $\boldsymbol{\xi}$ is a matrix, rather than a vector.

\subsubsection{Principles of Multilevel Summation}

The goal of the multilevel summation method is to approximate a dense product
$\mathbf{Gx}$ via a hierarchical expansion. For
molecular dynamics, the dense matrix 
$\mathbf{G}$ represents the interactions between particles.
To accelerate the computation, we replace it by a
sum of matrices which we associate with a hierarchy of increasingly coarse grids;
the entries of these matrices correspond to interactions between grid points. 
Finally, we interpolate the resulting products from the grids to continuous space,
so that they again correspond to interactions between particles.

\begin{figure}[ht]
  \begin{flushleft}
  {1.)}\hspace{4mm} \raisebox{-.5\height}{\includegraphics[height=9mm]{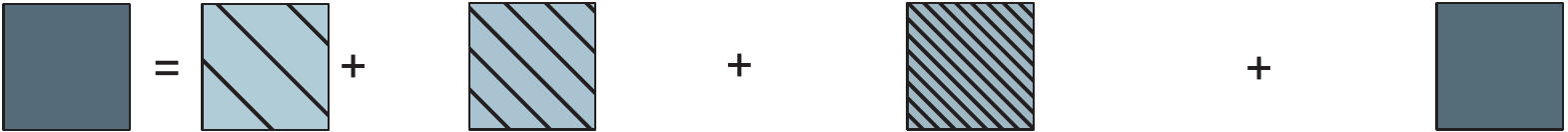}}\\
  {2.)}\hspace{4mm} \raisebox{-.5\height}{\includegraphics[height=9mm]{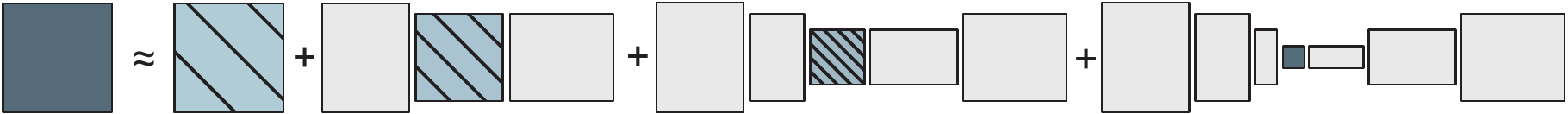}}
  \end{flushleft}
  \caption{The two major algorithmic steps of multilevel summation. 
    First, $\mathbf{G}$ is split into a sequence of increasingly denser
    matrices. These matrices are then approximated
    by restricting them onto smaller vector spaces,
    each of which corresponds to an increasingly coarse grid.
    To turn this method into a hierarchical calculation, we expand each term 
    into a sequence of nested matrix products. 
    The hashing
    denotes relative matrix density.}
  \label{figure:outerproduct}
\end{figure}

\autoref{figure:outerproduct} illustrates these basic concepts. 
The first line shows the initial decomposition of $\mathbf{G}$, which allows us to
make the calculations local on most grids.
In the second line, each term is expanded into a product through left and
right multiplication by interpolation matrices. 
To make the representation recursive,
we reuse the same interpolation matrices in a nested sequence for higher-order terms.

\definecolor{col1}{RGB}{53,53,53}
\definecolor{col2}{RGB}{128,128,128}
\definecolor{col3}{RGB}{204,204,204}
\definecolor{mat1}{RGB}{0,153,230}
\definecolor{mat2}{RGB}{51,191,230}
\definecolor{mat3}{RGB}{102,230,230}
\definecolor{mat4}{RGB}{153,230,215}
\definecolor{mat1}{RGB}{0,115,230}
\definecolor{mat2}{RGB}{46,168,230}
\definecolor{mat3}{RGB}{92,230,229}
\definecolor{mat4}{RGB}{153,230,215}
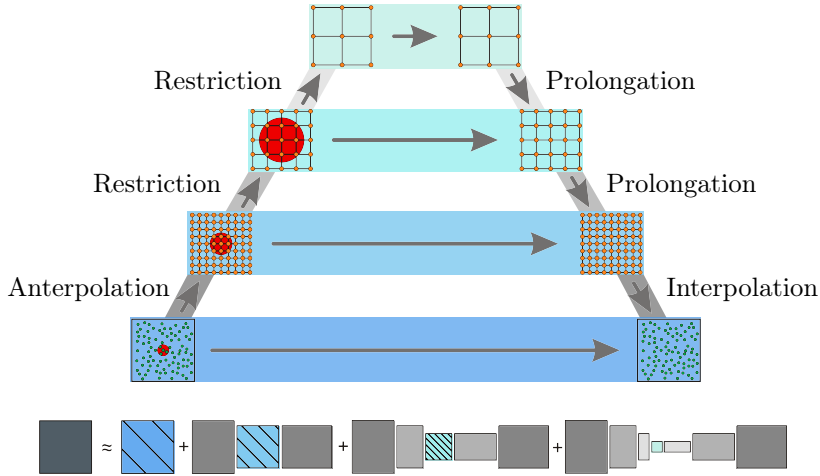
\begin{figure}[ht]
  \begin{center}
    \begin{tikzpicture}
      \fill [fill=col1, opacity=0.5] (-3.3,0.91) -- (-3.0,1.48) -- (-2.65,1.48) -- (-2.95,0.91) -- cycle;
      \fill [fill=col1, opacity=0.5] (3.3,0.91) -- (3.0,1.48) -- (2.65,1.48) -- (2.95,0.91) -- cycle;
      \fill [fill=col2, opacity=0.5] (-2.5,2.32) -- (-2.2,2.84) -- (-1.85,2.84) -- (-2.15,2.32) -- cycle;
      \fill [fill=col2, opacity=0.5] (2.5,2.32) -- (2.2,2.84) -- (1.85,2.84) -- (2.15,2.32) -- cycle;
      \fill [fill=col3, opacity=0.5] (-1.7,3.67) -- (-1.4,4.21) -- (-1.05,4.21) -- (-1.35,3.67) -- cycle;
      \fill [fill=col3, opacity=0.5] (1.7,3.67) -- (1.4,4.21) -- (1.05,4.21) -- (1.35,3.67) -- cycle;

      \fill [fill=mat1, opacity=.5] (-3.75,0.06) rectangle (3.75,0.91);
      \fill [fill=mat2, opacity=.5] (-3.0,1.48) rectangle (3.0,2.32);
      \fill [fill=mat3, opacity=.5] (-2.2,2.84) rectangle (2.2,3.67);
      \fill [fill=mat4, opacity=.5] (-1.4,4.21) rectangle (1.4,5.06);
      \node [anchor=base] at (0,0.05) {\includegraphics[width=0.46\textwidth]{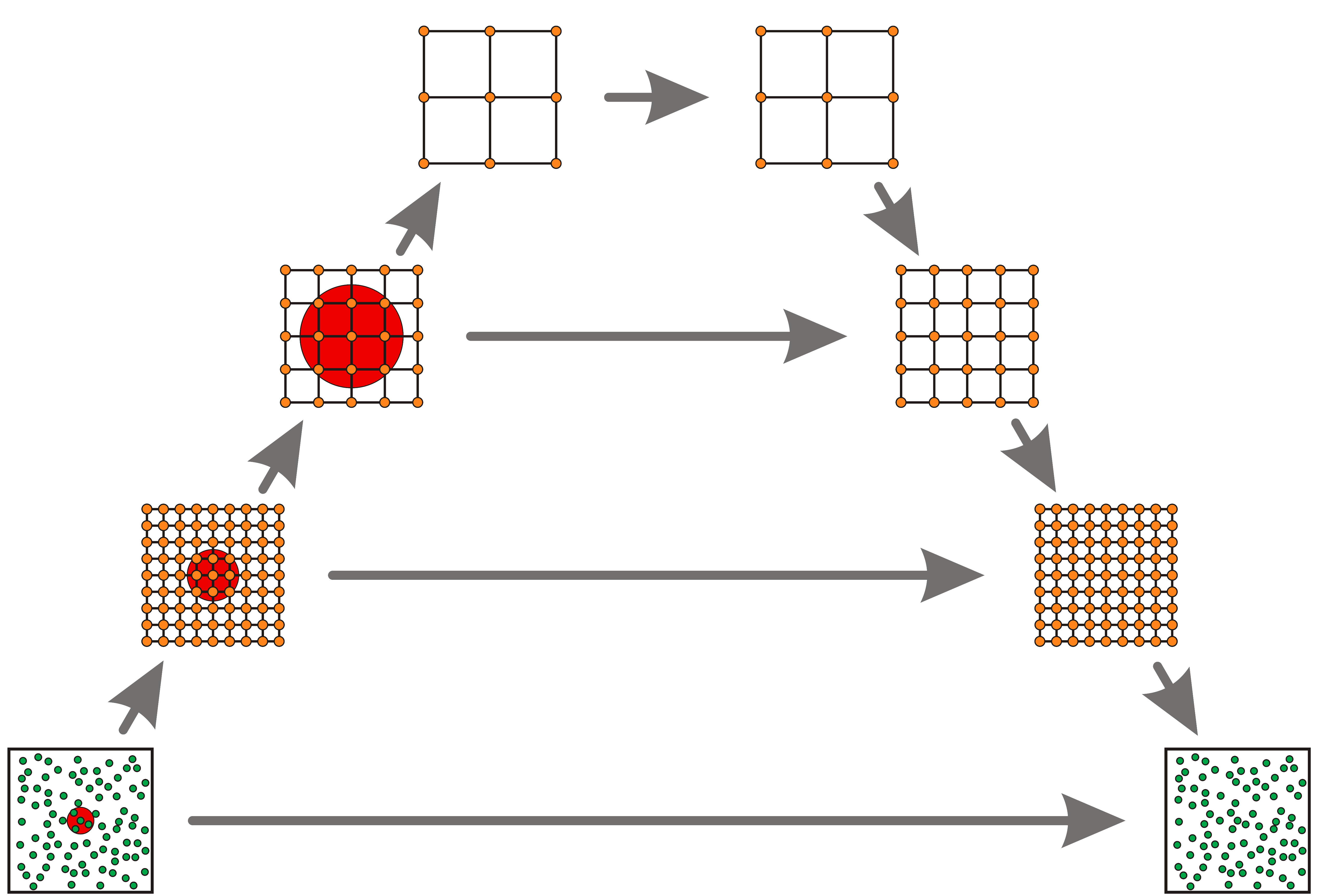}};
      \node [anchor=base, minimum height=16pt] at (-4.3,1.2) {\footnotesize{Anterpolation}};
      \node [anchor=base, minimum height=16pt] at (4.3,1.2) {\footnotesize{Interpolation}};
      \node [anchor=base, minimum height=16pt] at (-3.4,2.58) {\footnotesize{Restriction}};
      \node [anchor=base, minimum height=16pt] at (-2.6,3.94) {\footnotesize{Restriction}};
      \node [anchor=base, minimum height=16pt] at (3.5,2.58) {\footnotesize{Prolongation}};
      \node [anchor=base, minimum height=16pt] at (2.7,3.94) {\footnotesize{Prolongation}};
    \end{tikzpicture}
    \\[2mm]
    \begin{tikzpicture}[scale=1.3]
      % matrices
      \fill [fill=mat1, opacity=.6] (-2.953,-0.001) rectangle (-2.415,0.537);
      \fill [fill=mat2, opacity=.6] (-1.785,0.054) rectangle (-1.354,0.487);
      \fill [fill=mat3, opacity=.6] (0.123,0.135) rectangle (0.398,0.409);
      \fill [fill=mat4, opacity=.6] (2.41,0.213) rectangle (2.527,0.330);
      % interpolation operators
      % 1
      \fill [fill=col1, opacity=.6] (-2.237,0.001) rectangle (-1.805,0.537);
      \fill [fill=col1, opacity=.6] (-1.328,0.057) rectangle (-0.817,0.486);
      \fill [fill=col1, opacity=.6] (-0.621,0.003) rectangle (-0.187,0.539);
      \fill [fill=col1, opacity=.6] (0.866,0.057) rectangle (1.375,0.486);
      \fill [fill=col1, opacity=.6] (1.536,0.003) rectangle (1.969,0.539);
      \fill [fill=col1, opacity=.6] (3.275,0.057) rectangle (3.785,0.486);
      % % 2
      \fill [fill=col2, opacity=.6] (-0.167,0.057) rectangle (0.105,0.486);
      \fill [fill=col2, opacity=.6] (0.415,0.135) rectangle (0.847,0.409);
      \fill [fill=col2, opacity=.6] (1.988,0.0575) rectangle (2.261,0.486);
      \fill [fill=col2, opacity=.6] (2.825,0.135) rectangle (3.258,0.409);
      % % 3
      \fill [fill=col3, opacity=.6] (2.277,0.1345) rectangle (2.392,0.409);
      \fill [fill=col3, opacity=.6] (2.538,0.213) rectangle (2.812,0.330);
      % picture
      \node [anchor=base] at (0,0) {\includegraphics[width=0.598\textwidth]{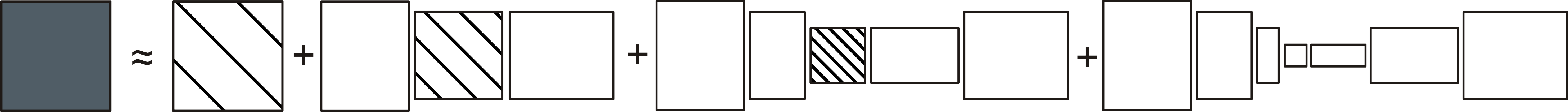}};
    \end{tikzpicture}
  \end{center}
  \caption{Schematic representation of one MLS step with three steps:
    1) the particles are mapped to the finest grid and from grid to grid;
    2) the corresponding potentials are evaluated, where a growing cutoff occurs;
    3) the results are interpolated back to the particles.}
  \label{figure:algorithm}
\end{figure}

%We distribute the original matrix so that
%interactions on all grids except the coarsest one are local,
%implying that the corresponding matrices are sparse. As the grids
%become coarser, the locality requirement is fulfilled
%by allowing the interaction cutoff to grow from grid to grid.

In practice, the matrices are not actually interpolated;
instead, the associativity of matrix-matrix
products is exploited, as shown in \autoref{figure:algorithm}.
First, we map the vector $\mathbf{x}$ to the finest grid in an anterpolation step.
Next, we recursively restrict the resulting vector to coarser grids before evaluating the potential
 on the grids. We then prolongate the results recursively to the
finest grid, and interpolate from the finest grid to the particles.
For the coarsening of the grids in the hierarchy, the grid spacing is
increased by a factor of two in each direction. %  This strategy ensures
% that successive grids share as many points as possible, which is
% advantageous for the restriction and the prolongation steps.

% Although the general outline and terminology of multilevel
% summation is similar to the multigrid V-cycle,\cite{Trottenberg2000}
% there are important differences.
% Unlike multigrid, the MLS method is non-iterative:
% calculations are performed only once per grid per computational
% cycle.  Moreover, all grids contribute to the final result; in a
% multigrid algorithm, solutions from coarser grids are instead used only as
% starting points for calculations on a finer grid after
% prolongation.

\subsubsection{Local and grid-based components of the potential}

The MLS method cannot be directly applied to functions such as
$r^{-6}$ that are only asymptotically smooth,\cite{Brandt1990} unless
the kernel is split into a smooth term and a remainder term.  The
smooth part can be calculated using MLS, while the asymptotically smooth
remainder term decays much faster than the
original, so that its computation will be as fast as the grid-based calculations.\cite{Brandt1990}

A similar approach is adopted in Ewald-like methods, as singularities
are disadvantageous in Fourier space. The potential is split into a
 singular term that decays much
faster than the original, so that a cutoff method can be used, and a smooth term
evaluated in Fourier space.
Ewald methods use exponentials and error functions to split the
kernel.
While advantageous for the Fourier space calculations, they are expensive for the real-space calculations.
By comparison, as it does not use Fourier transforms, the MLS method can rely on
polynomials, which are cheap to evaluate in real space.
More importantly, when a cutoff is used, exponentials and error
functions are truncated, thus introducing an error in Ewald
methods; in the MLS method, the asymptotically smooth part
is identically zero beyond a cutoff parameter $a$.

Another property MLS shares with mesh-based Ewald methods is that
 particles experience ``self-interactions'' on the grids.
Fortunately, this unphysical effect is easily corrected by means of a \emph{self-interaction energy}.
Since these self-interactions are time-independent, they can be calculated
once at the beginning of the simulation, as in the PPPM method.\cite{Toukmaji1996}

We can therefore write the matrix $\mathbf{G}$ as
\begin{equation}
  \mathbf{G} = \mathbf{G}^{(0)} + \tilde{\mathbf{G}} - c_{\textrm{self}} \, \mathbf{I};
  \label{eq:initial-splitting}
\end{equation}
where $\mathbf{G}^{(0)}$ is a sparse matrix containing the singularities,
$\tilde{\mathbf{G}}$ includes the smooth part of the kernel, to which
the MLS is applied,
and the last term is the self-interaction term.
The corresponding potential $V$ is
\begin{equation}
  V = \boldsymbol{\xi}^T \mathbf{G}^{(0)} \boldsymbol{\xi}
  + \boldsymbol{\xi}^T \tilde{\mathbf{G}} \boldsymbol{\xi}
  - V_{\textrm{self}}.
  \label{eq:pot-initial-splitting}
\end{equation}
The first two terms on the right-hand side of \autoref{eq:pot-initial-splitting}
are usually referred to as the short-range and long-range parts, respectively.\cite{Hardy2006, Hardy2009}
However, as the long-range part actually involves several short-range terms that become progressively longer-ranged, with the last term being purely long-ranged, we denote the distinctions as \emph{local} and \emph{grid-based}.

We introduce a ``smoothing function'' $\gamma$ to split $r^{-6}$ into local and grid-based parts,
\begin{equation}
  f(r) = \frac{1}{r^6}-\frac{1}{a^6}\gamma\left(\frac{r}{a}\right)+\frac{1}{a^6}\gamma\left(\frac{r}{a}\right).
  \label{eq:initial-splitting-f(r)}
\end{equation}
We require $\mathbf{G}^{(0)}$ to be sparse, and that its elements be
zero for all pairs of particles $(i,j)$ with $r_{ij} \ge a$.
Therefore, we define $\rho = r/a$ and choose $\gamma(\rho) =
\rho^{-6}$ for $\rho \ge 1$, so that the first two terms on the right-hand
side of \autoref{eq:initial-splitting-f(r)} cancel for $r\ge a$.
For $\rho < 1$, the choice is flexible, but $\gamma(\rho)$ should be continuous and continuously
differentiable to guarantee smooth potentials and forces.
It has been suggested that smoothing functions based on those used for the electrostatic $1/r$ potential can be used for $r^{-2}$ potentials;\cite{Stone2007}
we adopt this approach for the dispersion potential as well, leading to
functions such as
\begin{equation}
  \gamma(\rho)=\begin{cases} \frac{15}{8}-\frac{5}{4} \rho^{12} +\frac{3}{8} \rho^{24}, &\text{ for }
    \rho < 1, \\ \rho^{-6}, &\text{ for } \rho\ge 1.\end{cases}
  \label{eq:gamma}
\end{equation}
We do not claim that our choice of $\gamma(\rho)$ is optimal; further work is required to determine the ideal form.

The first two terms on the right-hand side of~\autoref{eq:initial-splitting-f(r)} cancel for $r>a$, so it can be calculated exactly
using a cutoff; as a consequence, errors only arise due to
approximating the last term on the grids. The last term provides also
the formulation for the self-interaction energy, because its
evaluation at $r=0$ gives the self-interaction of a particle
on a grid:
\begin{equation}
V_{\textrm{self}} = c_{\textrm{self}} \;\boldsymbol{\xi}^T \boldsymbol{\xi} = \frac{\gamma\left(0\right)}{a^6} \; \boldsymbol{\xi}^T \boldsymbol{\xi}.
\label{eq:self-interaction-energy}
\end{equation}

\subsubsection{Distribution of the non-local components}

\begin{figure}[ht]
  \begin{center}
    \includegraphics{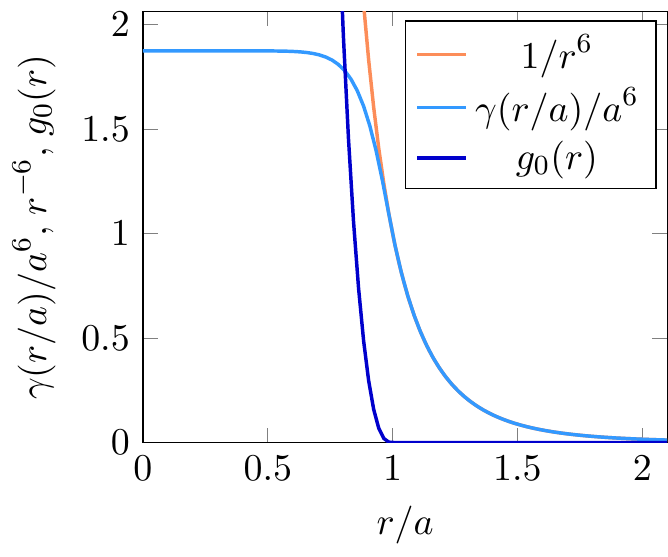}
    \includegraphics{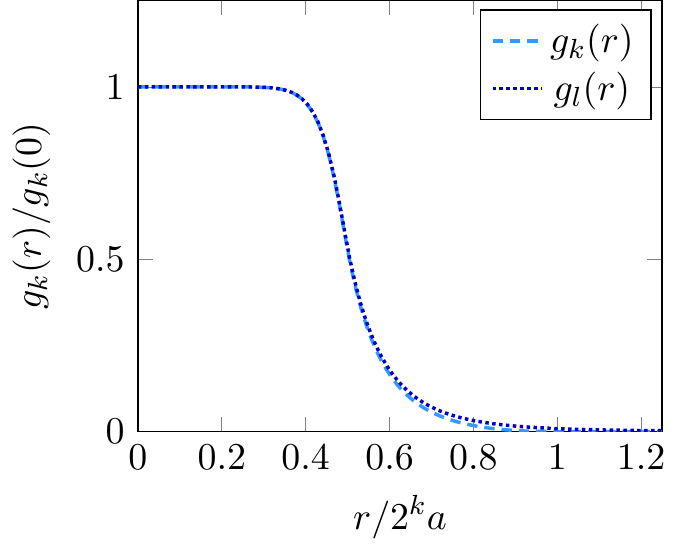}
    \caption{The smoothing function and $g_i$'s: (left) the original kernel $r^{-6}$ (red); the smoothing function of \autoref{eq:gamma} (blue); plus the difference of the former, the function $g_0$ (dark blue); (right) the normalized functions $g_k$, $1 \le k \le l$ (dark blue) and $g_l$ (blue).}
    \label{figure:smoothing-function-and-g_i}
    \end{center}
\end{figure}

  We approximate $\mathbf{G}$ on multiple grids by splitting $\tilde{\mathbf{G}}$
  into multiple terms:
  \begin{equation}
    \mathbf{G} = \mathbf{G}^{(0)} + \mathbf{G}^{(1)} + \cdots  + \mathbf{G}^{(l-1)} + \mathbf{G}^{(l)} - \frac{\gamma\left( 0 \right)}{a^6} \, \mathbf{I}.
    \label{eq:complete-splitting}
  \end{equation}
  Each grid, except the last, includes a cutoff:
  for all $\mathbf{G}^{(k)}$ with $k\ne l$, all entries with $r_{ij} \ge 2^k a$ must be zero.
  On the last grid, it is assumed that all grid points interact with one another.
  Splitting the kernel requires that the new terms
  vanish for $r_{ij} \ge 2^k a$, except on the coarsest grid.
  The entries of $\mathbf{G}^{(0)}$, which represent the local part,
  are given by
  \begin{equation}
    g_0(r) = \frac{1}{r^6}-\frac{1}{a^6}\gamma\left(\frac{r}{a}\right),
    \label{eq:g_0}
  \end{equation}
  thus fulfilling the locality constraint.
  To obtain the entries of $\mathbf{G}^{(k)}$ for $1 \le k \le l-1$,
  we iterate this process of adding and subtracting terms, and replace the cutoff $a$ with $2^ka$:
  \begin{equation}
    \label{eq:g_k}
    g_k(r) = \frac{1}{2^{6(k - 1)} a^6} \gamma\left(\frac{r}{2^{(k-1)}a} \right) - \frac{1}{2^{6k} a^6} \gamma\left(\frac{r}{2^k a} \right).
  \end{equation}
  The last matrix, evaluated on grid $l$, contains only the correction term for grid $l - 1$:
  \begin{equation}
    g_{l}(r) = \frac{1}{2^{6l-6}a^6}\gamma\left(\frac{r}{2^{l-1}a}\right).
    \label{eq:g_l}
  \end{equation}
  The behavior of $g_k(r)$
  using the smoothing function of \autoref{eq:gamma}
  is depicted in
  \autoref{figure:smoothing-function-and-g_i}. Since the $g_k(r)$
  become increasingly smooth, accurate results are obtained even as the
  grids become coarser.

Moreover, as the grid spacing grows by a factor of two, the smoothing functions in
\autoref{eq:g_k} are evaluated for the same arguments, regardless of the grid.
This also motivates the doubling of the cutoff as the grids coarsen.
During a simulation, the grids are fixed, so constant expressions, such as the
$g_k$'s and the mapping of operators between grids, can be pre-computed.

\subsubsection{Approximating the non-local components}

The matrices in \autoref{eq:complete-splitting} have size $N \times N$.
To reduce the algorithmic complexity, we need to work with smaller matrices
$\mathbf{G}^{(k)}$, $k=1,..,l$,
corresponding to each grid, which we can then interpolate into local space.
By exploiting the grid hierarchy, 
the matrices are interpolated onto the next finer grid,
where it is then added to the existing matrix. 
%, and then interpolating the combined results.
This process continues until local space is reached.

Formally, the interpolation is carried out by operators
$\mathcal{I}_k$.  Those operators are, in principle, independent of
the potential, and map their arguments to the next finer grid, or in
the case of $\mathcal{I}_1$, to continuous space.  The approximation
procedure is designed so that the interpolated matrices are accurate
representations of the original ones. 
The overall quality of the
approximation is determined by how well the interpolation 
approximates~\autoref{eq:pot-initial-splitting}.

Consequently, the kernel is approximated by
\begin{equation}
\frac{1}{r^6} \approx g_0 + \mathcal{I}_1\left[g_1 + \mathcal{I}_2\left[g_2 \cdots + \mathcal{I}_{l-1}\left[g_{l-1} + \mathcal{I}_l\left[g_l\right]\right] \cdots \right]\right].
\label{eq:final-equation-with-interpolation-operator}
\end{equation}
We define the interpolation operator acting on $g_k$ as
\begin{equation}
\mathcal{I}_k\left[g_k\right]\left(\mathbf{x}^{k-1}_i,\mathbf{x}^{k-1}_j\right) = \sum_\mu \sum_\nu \phi_\mu^k \left(\mathbf{x}^{k-1}_i\right) g_k\left(\mathbf{x}^k_\mu,\mathbf{x}^k_\nu\right)\phi_\nu^k\left(\mathbf{x}^{k-1}_j\right),
\label{eq:interpolation-operator}
\end{equation}
for $1 \le k \le l$, where $\mathbf{x}^i$ indicates a position on grid $i$ for $i > 1$ or in continuum space for $i = 0$.
In \autoref{eq:interpolation-operator}, $\phi_\mu^k$ is a nodal basis function defined as
\begin{equation}
\phi_\mu^k\left(\mathbf{x}^{k-1}_i\right)= \Phi\left(\frac{x^{k-1}-x_\mu^k}{2^k h_x} \right)\Phi\left(\frac{y^{k-1}-y_\mu^k}{2^k h_y} \right)\Phi\left(\frac{z^{k-1}-z_\mu^k}{2^k h_z} \right),
\label{eq:def-phi}
\end{equation}
where $\Phi$ denotes a dimensionless basis function, which we choose to be an Hermite interpolant, as used for the electrostatic potential.\cite{Skeel2002, Hardy2006} The cubic Hermite, for example, is
\begin{equation}
\Phi(\xi)=\begin{cases} (1-\left| \xi \right|)\left(1+\left| \xi \right| -\frac{3}{2} \xi^2 \right), &\text{ for }\left| \xi \right| \le 1, \\ -\frac{1}{2}(\left| \xi \right|-1) \left(2-\left| \xi \right| \right)^2, &\text{ for } 1 <\left|\xi \right| \le 2,\\ 0 &\text{ otherwise.} \end{cases}
\label{eq:Phi-Hermite}
\end{equation}
Like the smoothing function $\gamma$, we do not claim that this  choice
of Hermitian interpolants is optimal. The possible relationship between the smoothing and
interpolating functions is of special interest; we begin to examine this relationship in
\autoref{section: results}, but room for improvement remains.

\subsection{Algorithmic steps of the Multilevel Summation Method}
\label{s:algorithm}

\begin{figure}[ht]
  \begin{center}
    \includegraphics[width=0.45\textwidth]{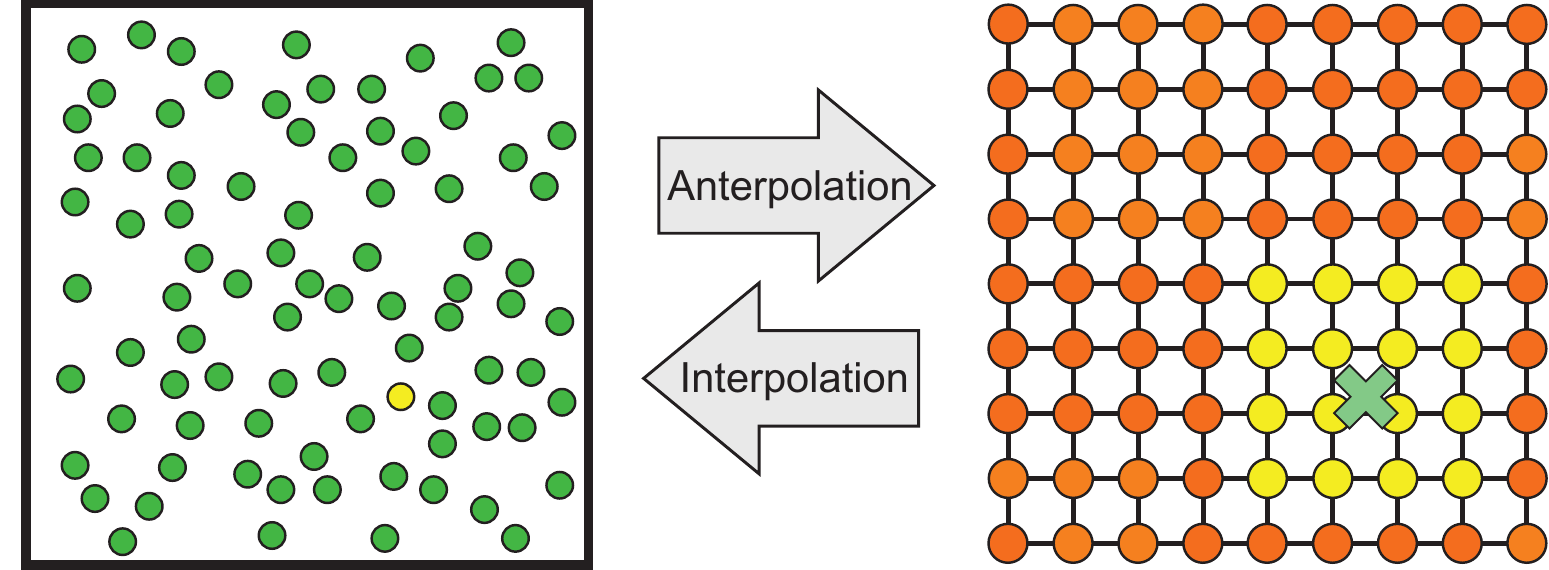}
    \hspace{2mm}
    \includegraphics[width=0.45\textwidth]{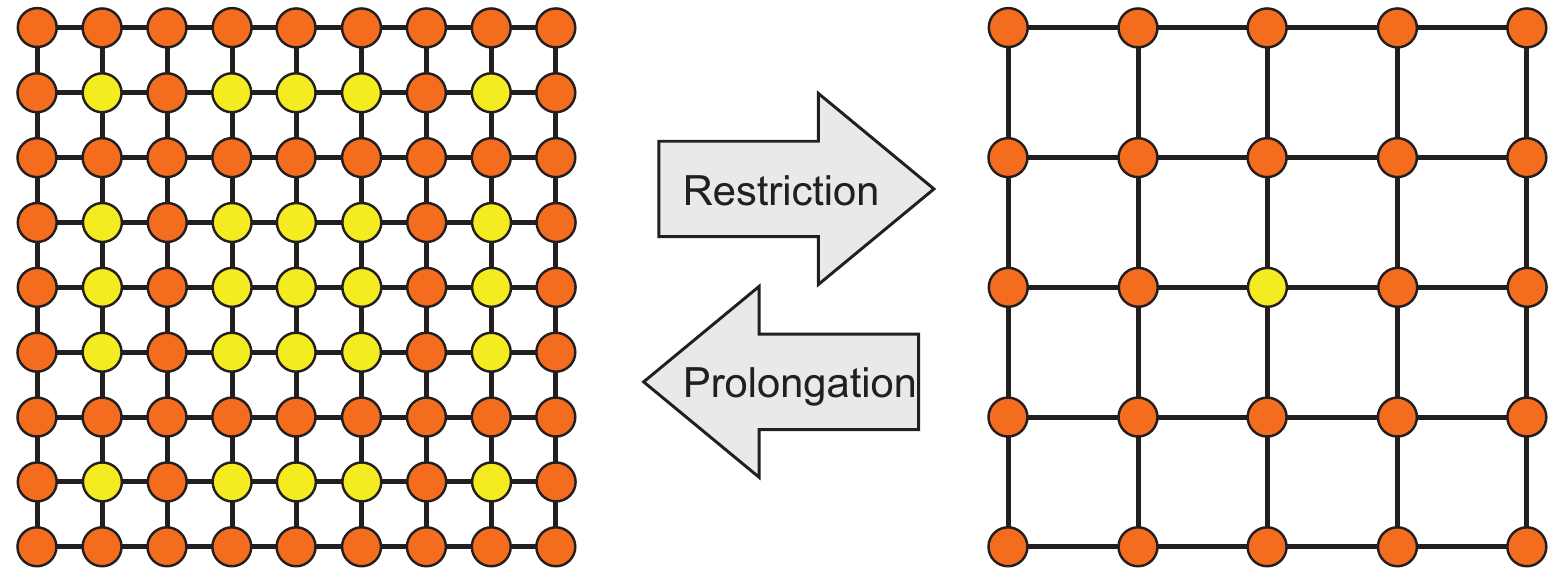}
    \vspace{2mm}\\
    \includegraphics[width=0.45\textwidth]{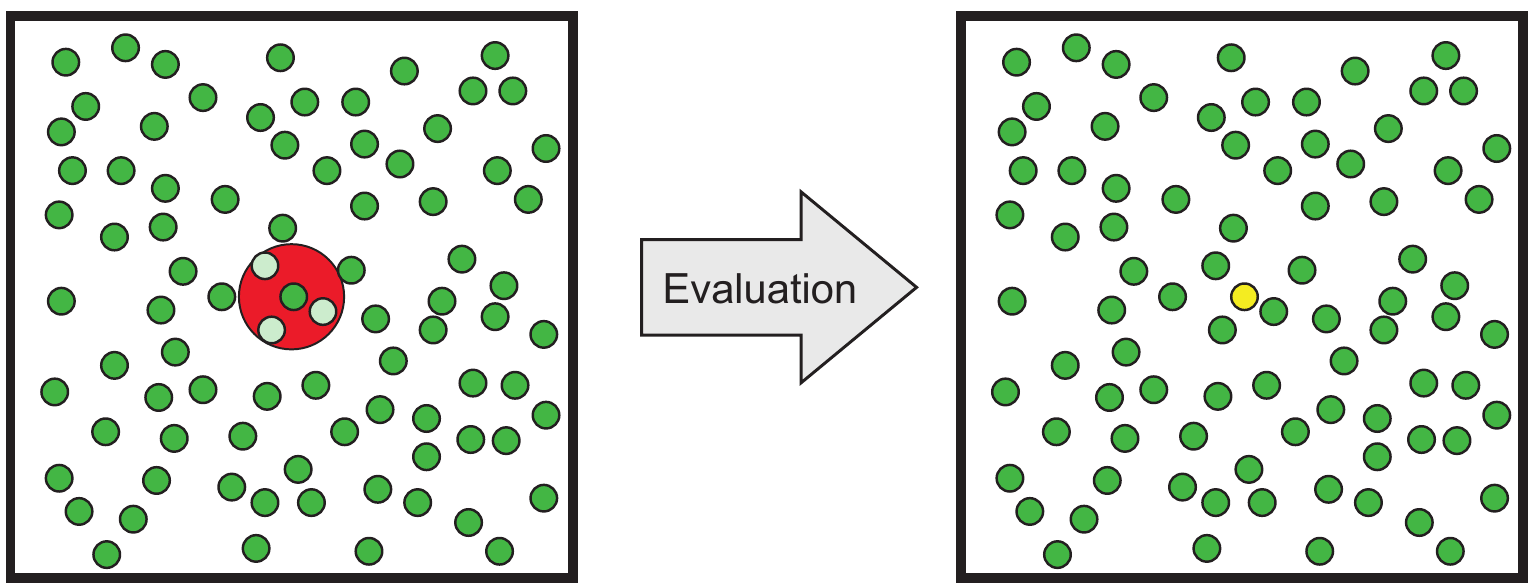}
    \hspace{2mm}
    \includegraphics[width=0.45\textwidth]{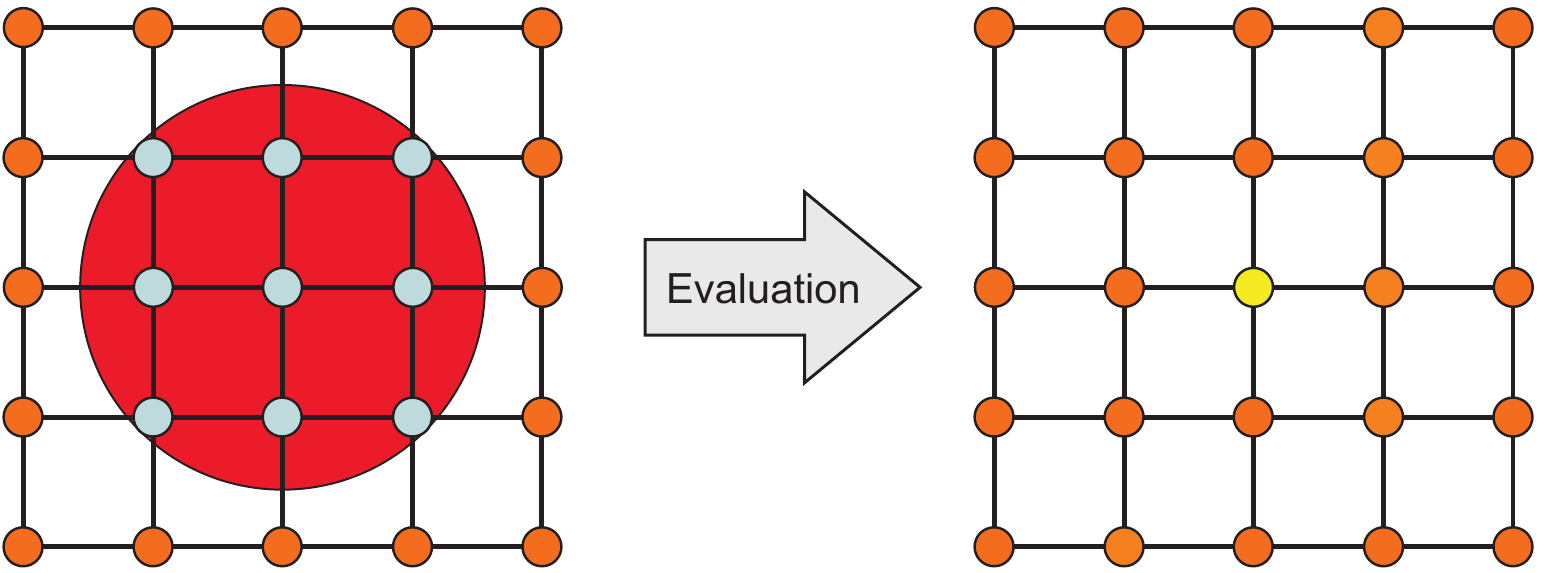}
  \end{center}
  \caption{Schematic representation of the various steps of the MLS.}
  \label{fig:steps-of-MLS}
\end{figure}

We present the following MLS algorithm in a potential-independent fashion. The steps are depicted in \autoref{fig:steps-of-MLS}, except for the evaluation on the coarsest grid, which is an all-to-all computation.

\begin{itemize}
   \item \emph{Anterpolation} approximates the vector $\boldsymbol{\xi}$ on the finest grid.
     We denote $\boldsymbol{\xi}$ as $\boldsymbol{\xi}^0$ to make the notation consistent: $\boldsymbol{\xi}^i$ is the respective approximation on grid $i$.
    The equation characterizing anterpolation is
    \begin{equation}
        \xi^1_\mu = \sum_j \phi_\mu^1(\mathbf{x}^0_j) \xi^0_j,
        \label{eq:anterpolation}
    \end{equation}
    where $\mu$ denotes a grid point and $j$ a particle.
    Because of the local support of the involved $\phi$'s, only grid points close to a particle are affected.

    \item \emph{Restriction}, as in multigrid methods, maps $\boldsymbol{\xi}^i$ from a grid $i$ to the next coarser grid $i+1$.
    Therefore, it is executed $l-1$ times, according to the expression
    \begin{equation}
        \xi^k_\mu = \sum_\nu \xi_\nu^{k-1} \phi_\mu^k(\mathbf{x}_\nu^{k-1}) \text{ for } k = 2,3,\dotsc,l.
    \end{equation}
    Like anterpolation, the local support of the involved $\phi$'s
    leads to a localization of the influence of any point on grid $i$ on the
    values on grid $i + 1$.

    \item The matrix-vector product is computed during the \emph{evaluation} step:
    \begin{equation}
        e_\mu^k = \sum_\nu g_{ka}^{LJ}(\mathbf{x}_\mu^k,\mathbf{x}_\nu^k)\xi_\nu^k \text{ for } k = 0,1,\dotsc,l.
        \label{eq:evaluation}
    \end{equation}
    For $k=0$, particles are directly considered, and for other values, the computations are performed on coarsened grids; for $k<l$, a cutoff is employed, while for $k=l$, an all-to-all computation on the last grid is performed. The evaluation step, as presented here, contains the short-range and the direct parts as defined by Hardy.\cite{Hardy2006}

    \item \emph{Prolongation}, as in multigrid methods, maps the vector $\mathbf{e}^i$ in \autoref{eq:evaluation} from grid $i$ to the next finer grid $i-1$, adding the term to the value of the vector on grid $i-1$ obtained from the evaluation step.
    Consequently, we perform $l - 1$ prolongation operations according to
    \begin{align}
        \mathbf{e}^{l,\text{tot}}_\mu &= \mathbf{e}^l_\mu\\
        \mathbf{e}^{k,\text{tot}}_\mu &= \mathbf{e}^k_\mu + \sum_\nu \phi_\nu^{k+1}(\mathbf{x}_\mu^k)\mathbf{e}^{k+1,\text{tot}}_\nu  \text{ for } k = l-1,\dotsc,2,1.
    \end{align}
    This process is the ``inverse'' of restriction.
    Accordingly, the local support of the $\phi$'s also leads to a
    localized effect of a point on grid $i$ on the values on grid $i - 1$; in
    practice, the same points on adjacent grids are involved in both
    restriction and prolongation.

    \item \emph{Interpolation} maps the vector $\mathbf{e}^{1,\text{tot}}$ back to the particles using
    \begin{equation}
        e_i^{0,\text{tot}} = \sum_\mu \phi_i^1(\mathbf{x}^0_i) e^{1,\text{tot}}_\mu.
        \label{eq:interpolation}
    \end{equation}
    This process is the ``inverse'' of the anterpolation.
    The same points to which we mapped $\boldsymbol{\xi}^0$ during anterpolation are now used to map $\mathbf{e}^{1,\text{tot}}$ back to the respective particle.
    Additionally, this step includes the scalar product to compute the potential:
    \begin{equation}
        V = \boldsymbol{\xi}^T \mathbf{e}^{0,\text{tot}}.
        \label{eq:interpolation-pot}
    \end{equation}
\end{itemize}

%%% Local Variables:
%%% mode: latex
%%% TeX-master: "DanielPaper01"
%%% End:

\subsection{Boundary Conditions}
\label{sec:last-grid-calc}

For non-periodic boundary conditions, target grid points can lie outside the domain, since anterpolation and restriction map objects to a region around initial particles or grid points.
Consequently, grids must extend beyond the actual domain boundaries, and span a larger physical space, as the grids coarsen.\cite{Brandt1990, Hardy2006}
However, the algorithmic and performance implications of grids growing beyond the original system size are negligible, and thus we only consider periodic boundary conditions.

Periodic boundary conditions complicate the computation of \autoref{eq:evaluation} on the coarsest grid, as there is no cutoff, and thus the calculation is nominally infinite. In the electrostatic case, periodic boundaries are trivial for charge-neutral systems: the requirement that the last grid consist of a single point leads to the grid's contribution to the energy vanishing.\cite{Hardy2006}  For dispersion potentials, however, $\boldsymbol{\xi}^0_i > 0$ and the term cannot vanish. Nevertheless, we can limit the work performed on the last grid, so that its computational cost is in principle the same as for non-periodic boundaries.

The origin of the problem with periodic boundary conditions becomes obvious
when we write down the equation used for the evaluation step on grid $l$, given in \autoref{eq:evaluation}:
\begin{equation}
  e_\mu^l = \sum_{\mathbf{i} \in\mathbb{Z}^3} \sum_{\nu=1}^{n_{gp}} g_l \left(r^{l\mathbf{i}}_{\mu \nu}\right) \xi^{l\mathbf{i}}_\nu,
  \label{eq:last-grid-periodic}
\end{equation}
where $\mathbf{i}$ is a vector of three indices identifying the image grid point, with the zero vector denoting the original image.
However, $\xi^{l\mathbf{i}}_\nu$ is the same for different values of $\mathbf{i}$, and we can thus omit this index.
Moreover, unlike the electrostatic potential, the summation is absolutely convergent, because the dispersion potential is short-ranged. Hence we can change the order of summation:
\begin{equation}
  \mathbf{e}_\mu^l = \sum_{\nu=1}^{n_{gp}} \left(\sum_{\mathbf{i} \in\mathbb{Z}^3}g_l \left(r^{l\mathbf{i}}_{\mu \nu}\right)\right) \boldsymbol{\xi}^{l}_\nu,
  \label{eq:last-grid-periodic-transformed}
\end{equation}
where the term in parentheses is time-independent and hence can be precalculated. Thus, computation of \autoref{eq:last-grid-periodic-transformed} is no more expensive than the non-periodic case.
Moreover, translational invariance implies that
if indices $\mu$ and $\nu$ are shifted by the same amount,
the sum does not change. Hence, we need to
precalculate only as many terms as there are points on the last grid.

\subsection{Mixing Rules}
\label{sec:mixing-rules}

To conclude our presentation of the MLS algorithm for dispersion, we define
the vector $\boldsymbol{\xi}$ introduced in
\autoref{eq:pot-vec-mat-vec}.
As mentioned, $\boldsymbol{\xi}$ is constant and its $i$th entry is a function of the properties of particle $i$.
The general form of the interparticle pair potential includes a repulsive term
as well as a dispersion term:
\begin{equation}
    V_{ij} = \frac{C_{ij}^{\textrm{rep}}}{r_{ij}^{12}} - \frac{C_{ij}^{\textrm{disp}}}{r_{ij}^6},
    \label{eq:disp-pair-pot}
\end{equation}
where $C_{ij}^{\textrm{rep}}$ and $C_{ij}^{\textrm{disp}}$ are determined by
the types of particles $i$ and $j$. In the remainder of the analysis, we ignore the repulsive term,
because for $r > \sigma$ it is negligible compared to the dispersion term. 
Compatiblility with \autoref{eq:pot-vec-mat-vec} requires
\begin{equation}
    V_{ij} = \frac{\xi_i \: \xi_j}{r_{ij}^6}.
    \label{eq:disp-pair-pot-multilevel}
\end{equation}
This is trivial if all particles are of the same type; for multiple species, the procedure and its complexity depend on the mixing rules.

For geometric mixing, we can choose
\begin{equation}
    \xi_i = \sqrt{C_{ii}},
    \label{eq:xi-general}
\end{equation}
where $C_{ii}$ denotes the interaction between particles of type $i$.
For the LJ potential, which contains the mixable parameters $\varepsilon$ and $\sigma$, the general formula for $C_{ij}$ is
\begin{equation}
    C_{ij} = 2 \varepsilon_{ij} \sigma_{ij}^6,
    \label{eq:xi-lj}
\end{equation}
so that
\begin{equation}
    \xi_i = \sqrt{2 \varepsilon_{ii}} \sigma_{ii}^3.
    \label{eq:xi-lj-geometric}
\end{equation}
By contrast, the Lorentz-Berthelot mixing rules use geometric mixing for $\varepsilon_{ij}$, but arithmetic mixing for $\sigma_{ij}$: $\sigma_{ij} = (\sigma_{ii} + \sigma_{jj})/2$.
To apply the MLS method, $\sigma_{ij}^6$ must be expanded as
\begin{equation}
V = \sum_{k=0}^6\boldsymbol{\xi}^T_k \mathbf{G} \boldsymbol{\xi}_{k-6},
\label{eq:disp-pair-pot-multilevel-LorentzBerthelot}
\end{equation}
where
\[
\xi_{i,k} = 
\frac{1}{8} \sigma_{ii}^k \sqrt{\binom{6}{k} 2 \varepsilon_{ii}}.
\]
For each of the seven terms in
\autoref{eq:disp-pair-pot-multilevel-LorentzBerthelot}, we can use
\autoref{eq:disp-pair-pot-multilevel}, as has been done for the dispersion
Ewald and PPPM methods.\cite{Isele-Holder2012,Veld:2007ip} This expansion is needed for
grid-based techniques because individual quantities, including polynomials
containing different powers of $\sigma_i$ and $\sigma_j$, must be handled
individually.
This change only affects performance, as seven instances of the MLS are used, compared to just one for geometric mixing rules.
Therefore, in \autoref{section: results} we consider only geometric mixing.

%%% Local Variables:
%%% mode: latex
%%% TeX-master: "DanielPaper01"
%%% End:

% Implementation

\section{Implementation of the Multilevel Summation Method}
\label{section: results}
\subsection{Implementation and Simulation Systems}
\label{section: implementation}

As a proof of concept, we created a C++ implementation of the MLS method for
dispersion. Since this code was meant as a prototype for future implementation
in open-source codes such as LAMMPS,\cite{Plimpton1995} we did not optimize parts
of the code, such as integrators, data structures, and the local potential
calculations, indirectly related to the MLS algorithm. Currently, our code
does not include support for polyatomic species, and is thus limited to
monatomic species, with support for geometric mixing rules.
All results presented are for serial runs on a single processor, and
performance measurements are based on timings for calculations over 1000 steps.

Our main test system was a LJ 12-6 gas containing 4000 particles in a slab of
dimensions $11.01\sigma \times 11.01\sigma \times 176.16\sigma$ with periodic
boundary conditions.  % Since $r^{-6}$ is calculated only for $r < a$, for
% efficiency the repulsive cutoff should be less than or equal to $a$. When the
% repulsive cutoff is equal to $a$, the case degenerates to adding an additional
% term to $g_0$.
A liquid slab was created by randomly placing LJ atoms within a
region comprising one-fourth of the total domain volume, followed by energy
minimization. To measure accuracy, we compared the potential energy and forces
for the initial configuration with a reference calculation in LAMMPS, using
the Ewald solver for dispersion\cite{Veld:2007ip} with a specified accuracy of
$10^{-12}$.

Most error measures for the forces are dominated by the quality of the
approximation of the largest forces in the system. However, the representation
of small forces can be relevant for the dynamics of the system. For example,
if several clusters of atoms are present, the forces between clusters will
be much smaller than the forces within a cluster, but the dynamics will
largely be influenced by the former. Instead, our error measure for the forces is
\begin{equation}
  \Delta F = \frac{1}{N} \sum_{i=1}^N \frac{ \| \mathbf{F}_i^{\text{MLS}} - \mathbf{F}_i^{\text{ref}} \|}{ \| \mathbf{F}_i^{\text{ref}} \| }.
\label{eq:error-measure-force}
\end{equation}
Although this measure becomes infinite when the force on a particle is zero, this is highly unlikely for MD systems.
The error in the potential energy is measured by
\begin{equation}
  \Delta E =  \frac{ | E^{\text{MLS}} - E^{\text{ref}} |}{ | E^{\text{ref}} | }.
\label{eq:error-measure-pot}
\end{equation}

\begin{figure}[ht]
  \center
  \includegraphics[width=0.789\textwidth]{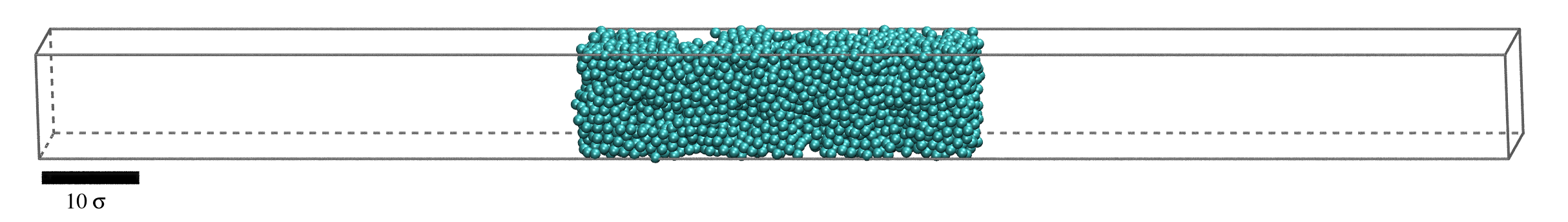}
  \hspace{8mm}
  \includegraphics[width=0.085\textwidth]{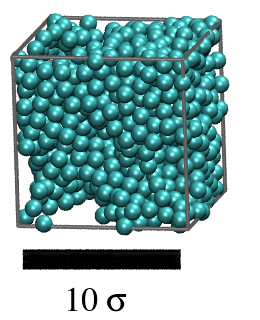}\\
  \includegraphics[width=0.78\textwidth]{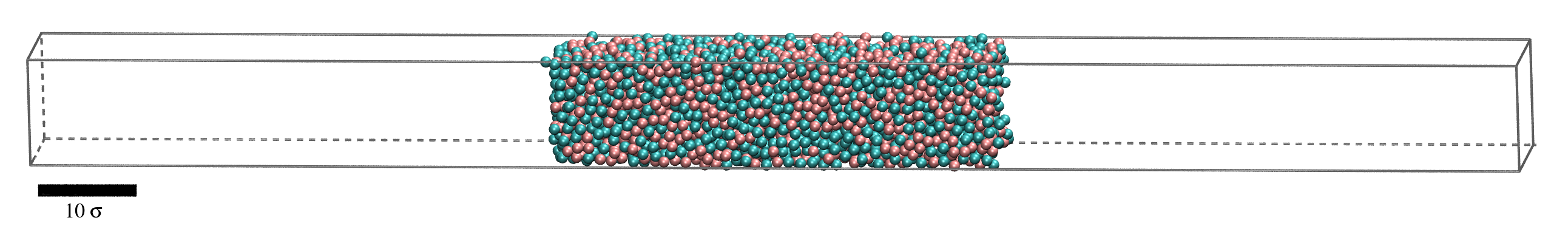}
  \hspace{8mm}
  \includegraphics[width=0.085\textwidth]{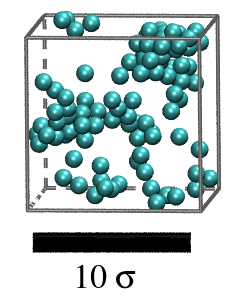}
\caption{Snapshot of the different simulation setups. (Top left) Slab geometry with 4000 particles. (Top right) Cubic geometry with 1000 particles. (Bottom left) Slab geometry with two different species. (Bottom right) Cubic geometry with dilute system of 100 particles.}
\label{figure: box}
\end{figure}
To test scalability, we also studied slab systems with 500, 32,000 and 256,000 particles, adjusting the domain size while preserving the aspect ratios. Furthermore, we investigated the performance of the MLS method using cubic domains of side length $11.01\sigma$ with 100 and 1000 particles. This test also allowed us to demonstrate that the performance is geometry-independent and scales well for dilute systems. Finally, we used a system consisting of two species, with differerent Lennard-Jones parameters, to illustrate that the accuracy and the performance are unaffected by mixing.
Illustrative sample starting configurations are depicted in \autoref{figure: box}. All grids used had uniform spacing; the same is assumed in our theoretical analyses below. Nevertheless, since restriction to uniform spacings is a matter of convenience, the results can be easily generalized to non-uniform grids.

\subsection{Error bounds}
\label{sec:error-bounds}

Following Hardy's derivation of strict error bounds for electrostatic
systems,\cite{Hardy2006} we can construct conservative bounds for
dispersion interactions. The derived bounds include constants for the
influence of the nodal basis function $\Phi$ and the smoothing
function $\gamma$, and it can be shown that the bounds for the
potential and for a component of the force vector are proportional to
$h^{n+1}/a^{n+7}$ and $h^{n}/a^{n+7}$, respectively. The value of $n$
is the largest integer less than or equal to $p-1$, where $p$ is the
interpolation order, for which the error bounds remain finite.  Finite
bounds are connected with the continuity of the smoothing function
$\gamma$: as in the electrostatic case,\cite{Hardy2006} if the
continuity is at least $C^{p-1}$, the highest possible effective order
is achieved.

\begin{figure}[ht]
  \begin{center}
    \includegraphics{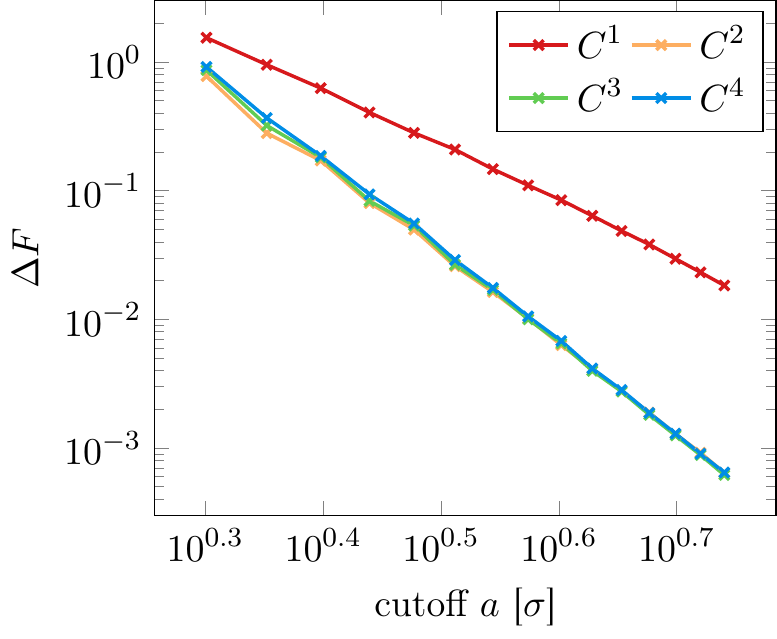}
    \includegraphics{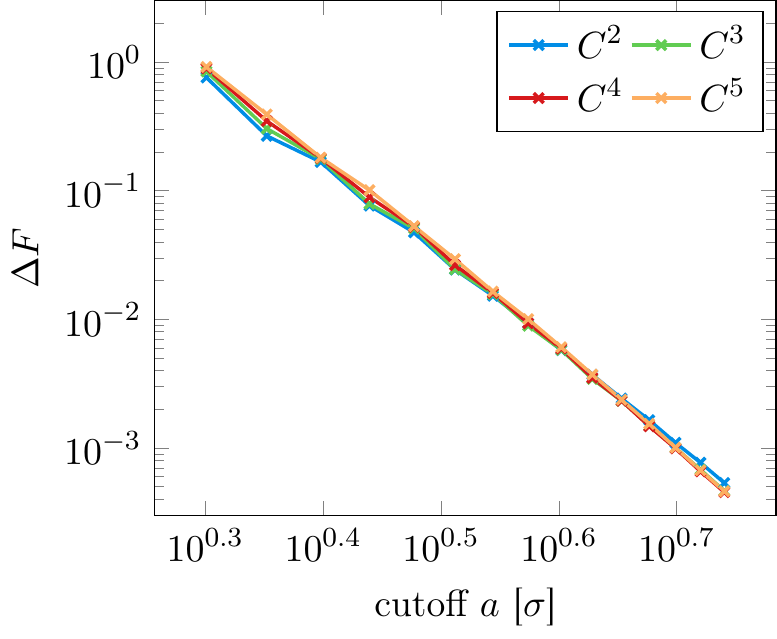}
    \caption{Error in the forces for the smoothing function $\gamma$ with interpolation order (left) $p=3$ and (right) $p=5$.}
    \label{figure:error-bounds-continuity}
  \end{center}
\end{figure}

For interpolation order $p=3$, this theoretical prediction agrees with numerical results. In the left-hand plot of \autoref{figure:error-bounds-continuity}, the different continuities for the smoothing function are applied to the 4000-particle system for interpolation order $p = 3$, as used in \autoref{eq:Phi-Hermite}. As the cutoff varies, the error for $C^1$ continuity decays more slowly than the other continuities. Moreover, continuity beyond $C^2$ does not improve the convergence rate, while the error increases slightly. This growth is due to increasingly large constants in the error bounds, supporting the theoretical prediction that $C^{p-1}$ continuity is optimal for interpolation order $p$. The argument for a $C^{p-1}$ smoothing function is that one wants the fastest possible convergence with the smallest possible error. However, in general this argument does not hold for dispersion interactions.

In fact, the advantage of a better convergence rate does not necessarily pay
off, as for small values of $a/h$, the importance of the error constants is
greater. In the right-hand plot of \autoref{figure:error-bounds-continuity} we
repeat the experiments for $p = 5$; for large values of $a/h$, the different
continuities for $\gamma$ are almost indistinguishable. Interestingly, for
small values of $a/h$, $C^2$ continuity has better accuracy than the others,
thus contradicting the theoretical prediction. Moreover, comparing the plots
for $p = 3$ and $p = 5$, we see that increasing the interpolation order does
not necessarily reduce the error. Thus, unlike the electrostatic case, we find
that optimal results are obtained for $C^2$ continuity with interpolation
order $p=3$, as higher continuities and interpolation orders increase the run
time of the program without significantly improving accuracy. We therefore use
these parameters for the rest of this paper, although
alternate choices of the smoothing and interpolation functions may 
influence these parameters.

\subsection{Factors influencing accuracy}
\label{sec:accuracy}

The ratio $a/h$ influences the error bounds: decreasing the spacing of the finest grid or increasing the cutoff yields more accurate results.
Numerical experiments were carried out to illustrate the convergence behavior with respect to both $a$ and $h$.

\begin{figure}[ht]
  \begin{center}
    \includegraphics{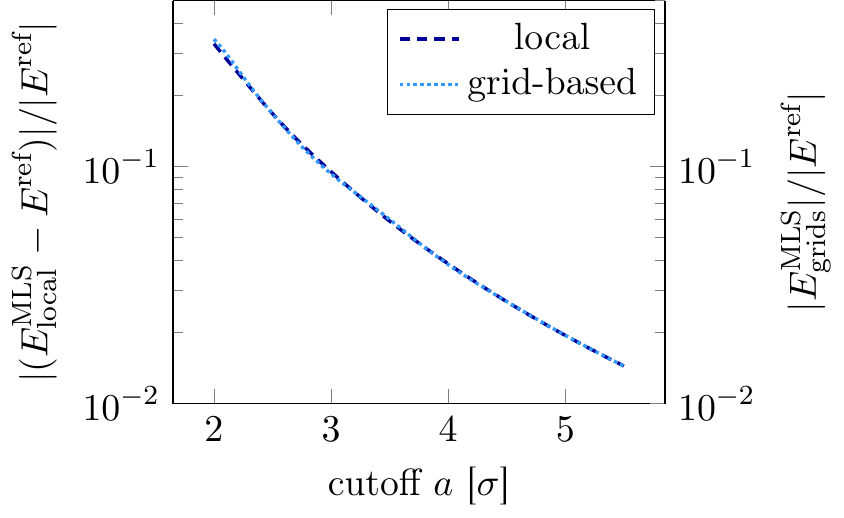}
    \includegraphics{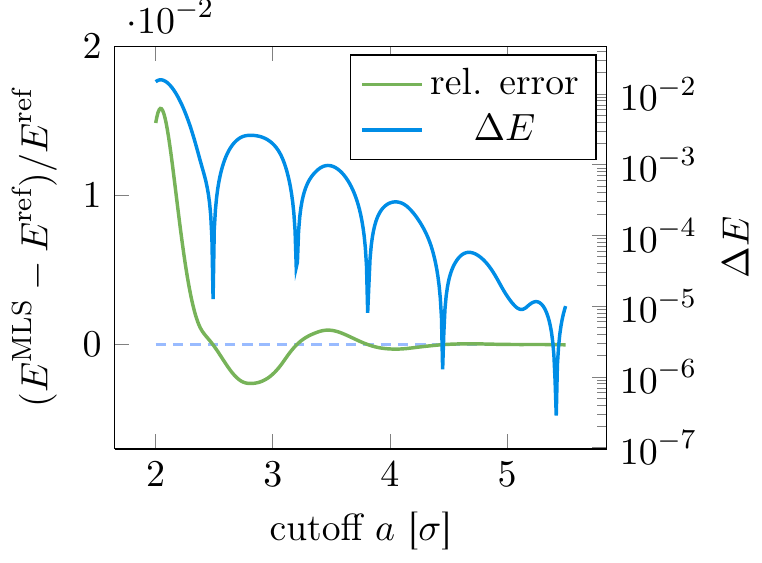}
    \caption{
      Contributions to the potential energy as a function of the
      cutoff. (Left) The absolute value of the relative difference between the
      local contribution to the potential and the reference potential (dark
      blue) nearly equals the absolute value of the relative grid-based contribution
      (light blue).  (Right) Relative difference between the overall potential
      energy calculated with the MLS method and the reference potential
      (green), together with its absolute value $\Delta E$ (dark blue).
    }
    \label{figure:short-long-range-potential-comparison}
  \end{center}
\end{figure}

We first examine the potential energy for different cutoffs. Results
for the 4000-particle system with finest-grid spacing of approximately
$1.38\sigma$ are presented in
\autoref{figure:short-long-range-potential-comparison}. The left-hand plot
shows that the local part converges to the exact solution as expected,
since as $a\rightarrow \infty$ the MLS becomes an $\mathcal{O}(N^2)$
method. We also verify that, in the same limit, the grid-based part
should converge to zero, as is also shown.
Moreover, the grid-based part approximates what is missing in the
local part for the correct solution; this is why the two
graphs in the left-hand plot are almost identical. The deviation of these plots is
the error of the MLS, which is caused solely by the approximations of
the grid-based part. As the grid-based part oscillates around the
local contribution, spikes in the error measure $\Delta E$ occur,
which are depicted in the right-hand plot.

\begin{figure}[ht]
  \begin{center}
    \includegraphics{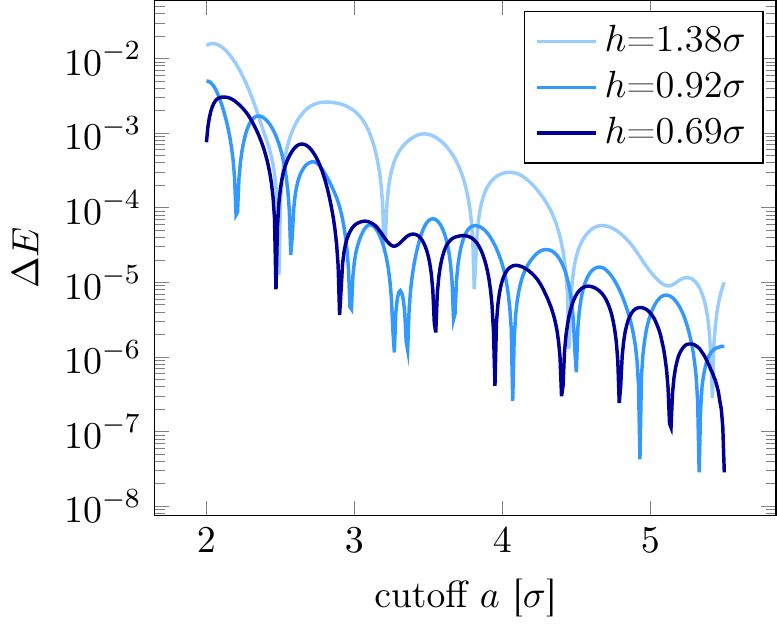}
    \includegraphics{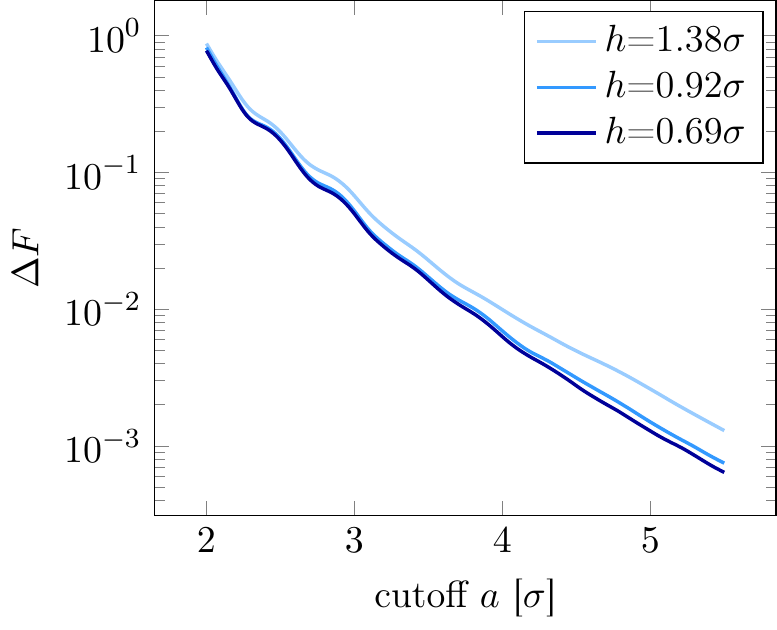}
    \caption{Error in the potential energy and in the forces as a function of
      cutoff; $h$ indicates the finest grid spacing.}
    \label{figure:different-spacing-different-cutoff}
  \end{center}
\end{figure}

To explore further the convergence of the total contributions to the potential
energy and forces, we examine the influence of the cutoff $a$ for three
finest-grid spacings $h$. In
\autoref{figure:different-spacing-different-cutoff} one can see that, as $a$
grows and $h$ decreases, the accuracy improves as expected. 
In addition, spikes in the
error of the potential energy occur for different values of $h$ rather than at
the same cutoffs.
Moreover, the
spacings $h=0.92\sigma$ and $h=0.68\sigma$ yield similar errors, while the
finer grid for $h=0.68\sigma$ consumes significantly more computational effort
(see \autoref{section: parameters run time}).  
We note that $h = 0.92\sigma$ corresponds to a grid with
$12\times12\times192$ points. For the Coulombic MLS
method with periodic boundary conditions, 
such a grid cannot be used, because the number of grid points in each
direction must always be a power of two. However, since this constraint does not apply to dispersion, grids like that with spacing $h = 0.92\sigma$ are possible.

Finally, we note that the last grid in the slab simulations never had fewer than sixteen points, as this would imply grid refinement in a single direction, which is impractical.

\begin{table}[ht] % values from plot_diff_spacings_same_ngrids.txt
  \caption{Influence of the finest-grid spacing on the error in potential
    energy and forces. The cutoff is $3\sigma$; three grid levels were
    used. }

  \begin{center}
    \begin{tabular}{C{0.13\linewidth}C{0.17\linewidth}C{0.17\linewidth}C{0.17\linewidth}C{0.17\linewidth}}
      \hline
      \hline
      $h$ [$\sigma$]& 2.75  & 1.38 & 0.69 & 0.34 \\
      \hline
      % EXACT: -7.0359090041738809873
      % E_total & -7.0444148780178120 & -7.0220459288857278 & -7.0355024049415604 & -7.0357905034612287 \\
      % $\Delta E$ & 0.0012089232306564904 & 0.0019703318050204024 & 5.7789154475897018e-05 & 1.6842274762506769e-05 \\
      $\Delta E$ & $1.21\times 10^{-3}$ & $1.97\times 10^{-3}$ & $5.78\times 10^{-5}$ & $1.68\times 10^{-5}$ \\
      % $\Delta F$ & 0.158306801483  & 0.0674925342399 & 0.0498439298008 & 0.0230155291421 \\
      $\Delta F$ & $1.58\times 10^{-1}$  & $6.75\times 10^{-2}$ & $4.98\times 10^{-2}$ & $2.30\times 10^{-2}$ \\
      \hline
      \hline
    \end{tabular}
  \end{center}
  \label{table:different-spacing-same-cutoff}
\end{table}

Since reducing the finest grid spacing $h$ quickly becomes
prohibitively expensive, in
\autoref{table:different-spacing-same-cutoff} we present results
for four different spacings, all with three grids, and a fixed cutoff of
$a = 3\sigma$. While one sees monotonic convergence in the
forces, this is not the case for potential energy. This can be
explained by the oscillatory convergence that led to the spikes in the
errors (such as in the right-hand plot in
\autoref{figure:short-long-range-potential-comparison}).

The number of grids used also affects the accuracy. The derived energy error
bounds for $l$ grids are a factor $1 + \mu^6 \left( 1 - 2^{-6\left(l-1\right)}
\right)/63$ larger than for a single grid, while the bounds for the force components are $1+
{\mu^6}\left( 1 - {2^{-7\left(l-1\right)}} \right)/127$ times larger when Hermite
interpolation functions are used, where $\mu$ is an error constant associated
with the nodal basis function $\Phi$.
Thus, increasing the number of grids always leads to larger bounds and thus to reduced accuracy. However, when $l$ is still relatively small, the effect of adding more grids is negligible. This can be seen in \autoref{table:error-bounds-more-grids}: the accuracies for $l = 3, 4$, and $5$ are nearly identical.

\begin{table}[th]
  \caption{Influence of the number of grids on the accuracy for the 4000-particle system with a finest-grid spacing of $0.688\sigma$.}
  \begin{center}
    \begin{tabular}{C{0.22\linewidth}C{0.2\linewidth}C{0.2\linewidth}C{0.2\linewidth}}
      \hline
      \hline
      \# grids $k$ & 3  & 4 & 5 \\
      \hline
      % E_total & -7.0355024049415604 & -7.0355022151167281 & -7.0355021591408287 \\
      % | (E_total_3 - E_total_k) / E_total_3 | & 5.7789154475897018e-05 & 5.7816133908431568e-05 & 5.7824089653657661e-05\\
      % $\displaystyle\dfrac{|\Delta E_3 -\Delta E_k|}{\Delta E_3}$ & 0 & 0.00046685979020167772 & 0.00060452827312456367\\
      \vspace{1mm}$\displaystyle\dfrac{|\Delta E_3 -\Delta E_k|}{\Delta E_3}\vspace{1mm}$ & 0 & $4.67\times 10^{-4}$ & $6.05\times 10^{-4}$ \\

      % $\Delta F$ & 0.0498439298008 & 0.0498438226163 & 0.0498438252282 \\
      % $\displaystyle\dfrac{\|\Delta F_3 -\Delta F_k\|}{\Delta F_3}$ & 0 & 2.1504022741819005e-06 & 2.0980007076867444e-06 \\
      \vspace{1mm}$\displaystyle\dfrac{\|\Delta F_3 -\Delta F_k\|}{\Delta F_3}$\vspace{1mm} & 0 & $2.15\times 10^{-6}$ & $2.10\times 10^{-6}$ \\
      \hline
      \hline
    \end{tabular}
  \end{center}
  \label{table:error-bounds-more-grids}
\end{table}

\begin{figure}[ht]
  \begin{center}
    \includegraphics{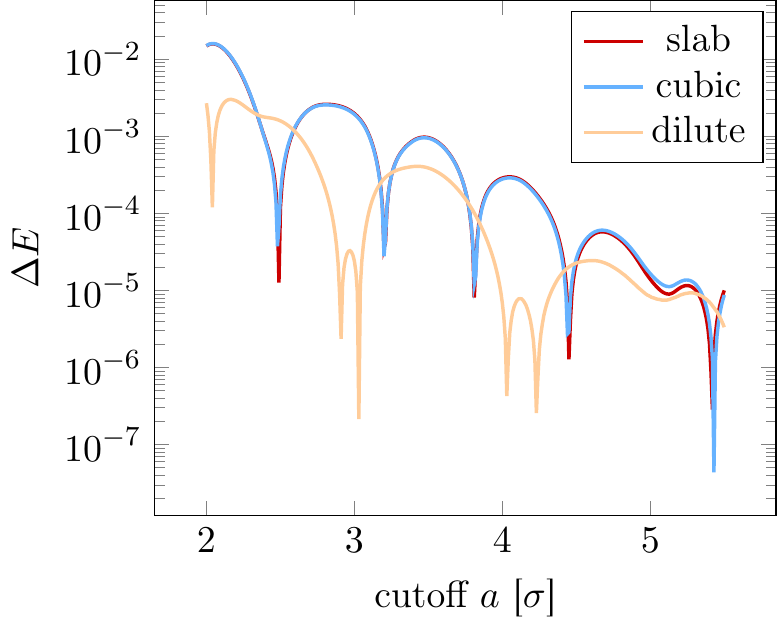}
    \includegraphics{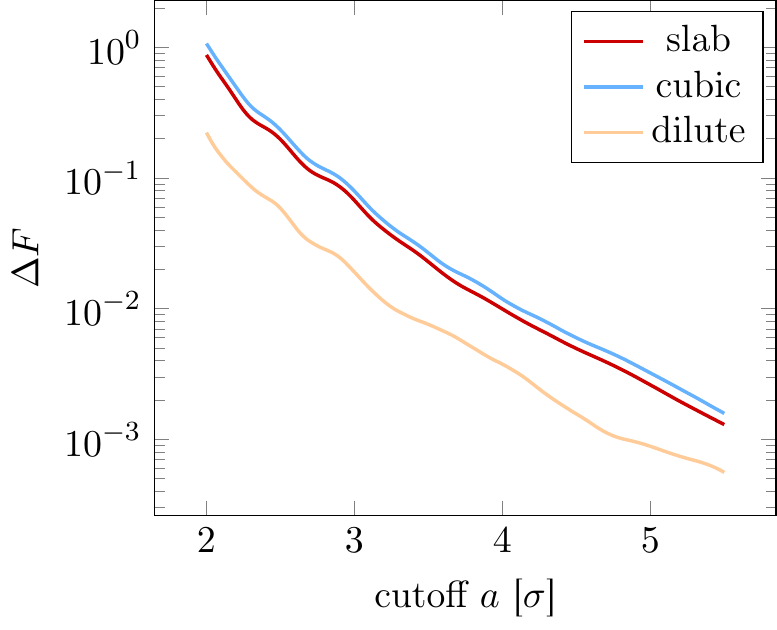}
    \caption{Error in potential energy and forces as a function of cutoff for slab, dense, and dilute cubic geometries.}
    \label{figure:different-systems-different-cutoff}
  \end{center}
\end{figure}

The impact of system size and geometry on the accuracy can be measured by examining the error for systems with different geometries. As seen in \autoref{figure:different-systems-different-cutoff}, changing from a slab to a cubic geometry has little effect on accuracy, while dilute systems tend to have greater accuracy than denser systems.

\begin{figure}[ht]
  \begin{center}
    \includegraphics{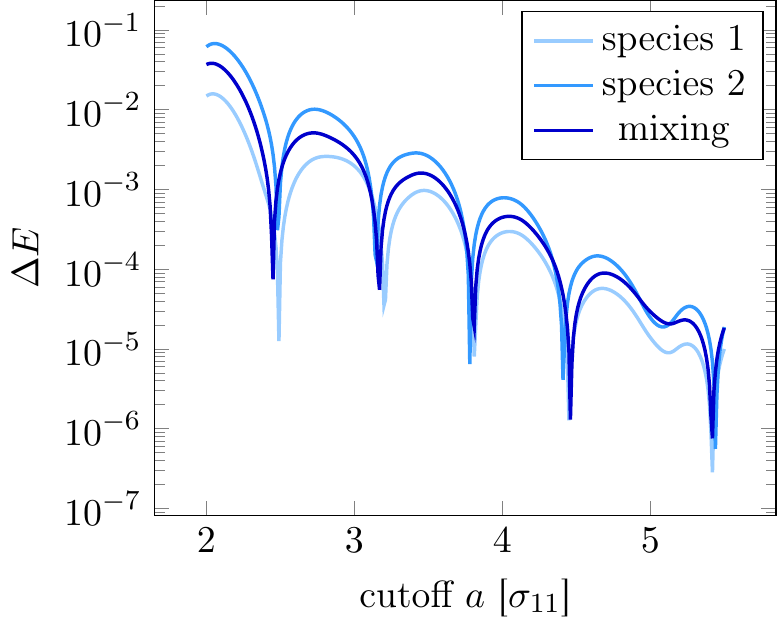}
    \includegraphics{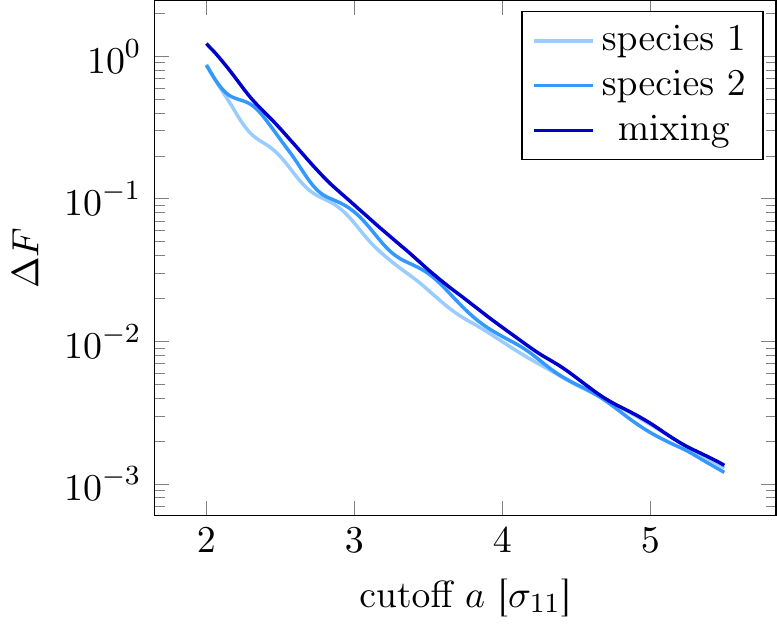}
    \caption{Error in potential energy and forces as a function of cutoff for two pure systems with different Lennard-Jones parameters and a mixture thereof. The cutoff $a$ is in units of $\sigma_{11}$.}
    \label{figure:mixing-accuracy}
  \end{center}
\end{figure}

Until now, we have only considered systems containing a single type of
particles. However, most MD simulations contain multiple particle
types. Therefore, we compared results for three 4000-particle systems with
identical experimental setups except for the system composition. The
comparison involves two pure systems and one system with two species in the
slab geometry above. The pure systems contain particles of species 1
($\sigma_{ii}=1$, $\varepsilon_{ii}=1$) and species 2 ($\sigma_{ii}=1.25$,
$\varepsilon_{ii}=0.5$), respectively. The mixed system contains particles in an equimolar mixture of both species, with geometric mixing rules used for $\sigma_{12}$ and $\varepsilon_{12}$.
The results in \autoref{figure:mixing-accuracy} show that there is basically no difference in accuracy between different setups. The cutoff is expressed in units of $\sigma_{11}$ since the accuracy is controlled by the ratio $a/h$, so that using different units for all three plots would skew the results.

\subsection{Factors influencing performance}
\label{section: parameters run time}

The run time is influenced by: the cutoff $a$; the finest-grid spacing $h$; the interpolation order $p$, which depends on the size of the support of the nodal basis function; the continuity of the splitting function~$\gamma$; and the number of grid levels $l$. As the qualitative impact on the run time should be independent of potential, the analysis will be similar to that for the electrostatic potential.\cite{Hardy2006}

The local part of the MLS method is a classical short-range
computation with cutoff $a$. The average work per particle to
calculate interactions inside the cutoff is
proportional to $(a/d)^3$, where $d = (V/N)^{1/3}$, the mean
nearest-neighbor distance in a homogeneous distribution. Because the work
per particle is independent of the number of particles, the total work
scales as $\mathcal{O}(N)$. As the other parameters are related to the
grid structure, the run time of the local part is additionally affected only by
the continuity of the splitting function~$\gamma$; however, its impact
is secondary and we do not investigate it further.

The dominant cost of the grid-based part occurs during the evaluation.
Except on the last grid, where no cutoff can be applied, the cost scales with the number of grid points times the effort per grid point.
The number of points $N_k$ on grid $k$ is equal to ${\left(2^{k-1}h\right)^{-3}}V$, where $V$ is the domain volume.
The work per grid point is proportional to the number of points inside the cutoff, which scales as $(a/h)^3$.
The effort per grid point is roughly constant between grids, as both the
spacing and the cutoff increase by a factor of two. 
Therefore, the overall cost for all cutoff calculations on grids is proportional to $(a/h)^3$ times the total number of grid points across all grids.
This total does not exceed $8/7$ of the number of points on the finest grid $N_1$.\footnote{$ \protect\sum_{k=1}^{l-1} N_k = \protect\sum_{k=1}^{l-1} \left(\frac{1}{8}\right)^{k-1} N_1 \le \protect\sum_{k=0}^\infty{\left(\frac{1}{8}\right)^k} N_1= \frac{N_1}{1-\frac{1}{8}} = \frac{8}{7}N_1$}
Consequently, the overall cost is proportional to $V a^3 / h^6$, most of which is associated with  the calculation on the finest grid.

The evaluation step on the last grid must be negligible compared to
the evaluations on lower grids for the method to remain linear in
complexity. However, even when the number of grids remains fixed, the
corresponding cost scales no worse than on the other grids: as
there is no cutoff, it scales as $N_l^2 = 2^{2l-2} h^{-6}$.

The influence of the other parameters on the cost of the grid-based part is
usually minor. Increasing the interpolation order leads to a larger support
for the nodal basis function, and thus a greater cost, as does increasing the
continuity of the splitting function $\gamma$, which leads to $\gamma$ having
a higher polynomial degree.
As shown in \autoref{figure:error-bounds-continuity}, though, for the basic choice of $\gamma$ used in the present work, increasing either the interpolation order or the continuity of $\gamma$ does not lead to significantly more accurate results.

\begin{figure}[ht]
  \begin{center}
    \includegraphics{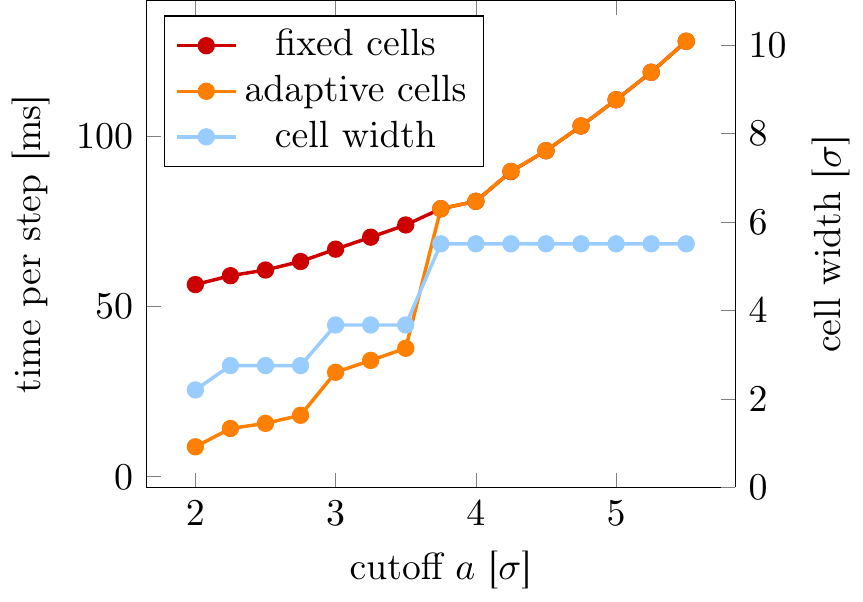}
    \includegraphics{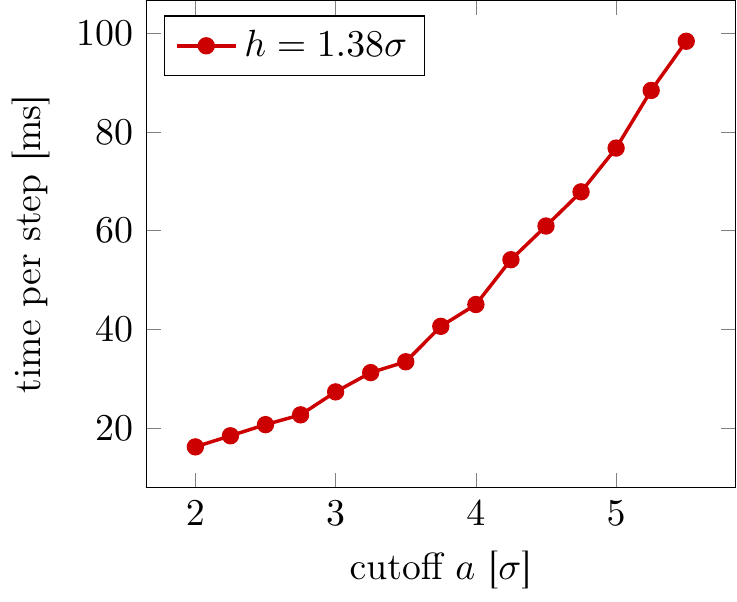}
    \caption{Timings for the local (left) and grid-based (right) part
      of our implementation. For the local timings with fixed cells,
      the same cell width was used for all simulations
      (5.505~$\sigma$), while with adaptive cells, the
      cell width was determined by the smallest value bigger than $a$
      for which the domain length divided by it remained an
      integer.
    }
    \label{figure:short-long-range-timings}
  \end{center}
\end{figure}

The timings for our prototype implementation for the 4000-particle
system in \autoref{figure:short-long-range-timings} reveal that in the
local part the cost of determining which particles interact can be
significant, while calculating the interactions is proportional to
$a^3$. In the linked-cell method, all particles that interact with a
given particle lie either the same cell or in neighboring
cells,\cite{Griebel2007} so using a fixed cell width for all cutoffs
reveals the cost of determining the interactions. In fact, the run
time in the left-hand plot does not converge to zero for $a\rightarrow
0$, although it shows the expected cubic behavior in $a$. In practice,
however, one chooses the cell width as close to $a$ as possible; the
cost for finding interactions then scales roughly as $a^3$, but
discontinuities in the timings occur whenever the cell width
changes. In the future, our implementation of the local part will
combine the linked-cell method with Verlet neighbor lists, thus
reducing the occurring discontinuities. As the grids do not change
during the simulation, the cost for finding grid points that interact
is minimal, resulting in a smooth, cubic scaling in $a$ for the
grid-based part.

\begin{figure}[ht]
  \begin{center}
    \includegraphics{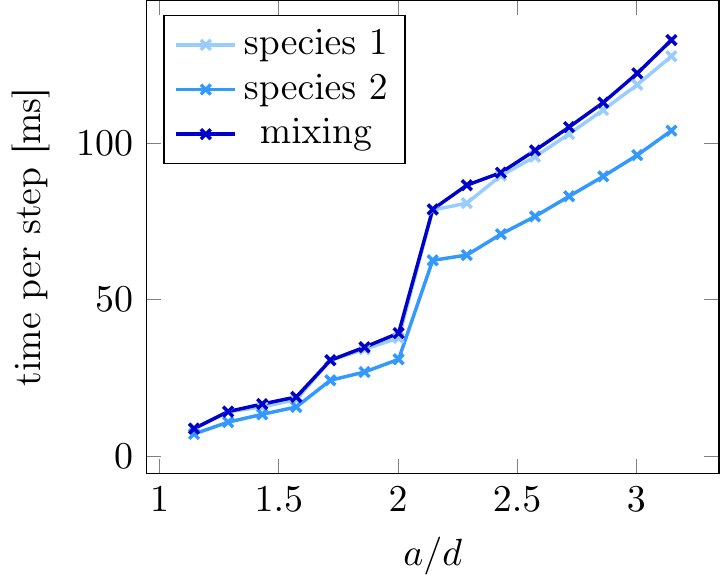}
    \includegraphics{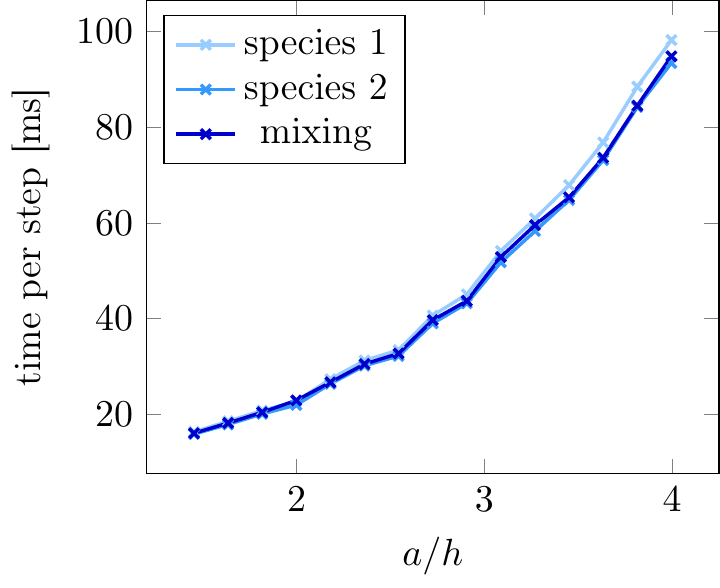}
    \caption{Timings of the local (left) parts with adaptive cell sizes and of the grid-based (right) parts for simulations with different mixing parameters.}
    \label{figure:mixing-timings}
  \end{center}
\end{figure}

In \autoref{figure:mixing-timings}
we show the performance for systems with multiple particle types.
To compare the timings fairly, the cutoff is divided by the characteristic length.
The larger deviation of species 2 compared to the other systems can be explained by the stronger repulsion associated with larger values of $\sigma$, so that a smaller number of particles interact at the beginning of a simulation. Since the same grids are used, we find no significant differences in the grid-based timings.

%%% Local Variables:
%%% mode: latex
%%% TeX-master: "DanielPaper01"
%%% End:

% Parameter Selection & Complexity

\subsection{Parameter selection for the Multilevel Summation algorithm}
\label{section:optimal-parameters}

For a given accuracy, we can determine optimal choices for $a$ and
$h$, assuming particles are homogeneously distributed. The error
bounds of \autoref{sec:error-bounds} imply that the absolute error in
the forces is proportional to $h^n/a^{n+7}$. As for dispersion a
smoothing function with $C^{p-1}$ is not always optimal, we do not replace $n$ with $p-1$, as done in the electrostatic case.\cite{Hardy2006} Instead, $n$ corresponds to the continuity of the smoothing function used.
Since the forces are proportional to $r^{-7}$ and the mean nearest neighbor distance between particles is $d$, the maximum magnitude of the total force is proportional to $d^{-7}$. Combining these observations, we obtain the following relation for the accuracy $\epsilon$:
\begin{equation}
	\epsilon = \frac{Ch^{n}d^7}{a^{n+7}},
	\label{eq:accuracy-tolerance}
\end{equation}
where $C$ is a constant. This equation implies that when the optimal finest-grid spacing is known, the cutoff can be calculated as
\begin{equation}
	a = \left( {\frac{Cd^7h^n}{\epsilon}} \right)^{\frac{1}{n+7}}.
	\label{eq:optimal-cutoff}
\end{equation}

To find the optimal spacing $h$, we note again that the computational cost is dominated by the evaluation step.
Although both the cost of the last-grid evaluation and the cost of restrictions and prolongations depend on the finest-grid spacing, we omit them because they are negligible compared to the overall cost, and because so doing makes it possible to find an analytic solution.\cite{Hardy2006}
Thus the dominant cost $c$, which depends on $a$ or $h$, can be modeled by
\begin{equation}
	c(a,h) = C_{\text{local}}\left(\frac{a}{d}\right)^3 N+C_{\text{grids}}\left(\frac{a}{h}\right)^3 \frac{V}{h^3},
	\label{eq:cost}
\end{equation}
where $C_{\text{local}}$ and $C_{\text{grids}}$ are constants describing the work done for the local and grid-based parts, respectively. Expressing $V$ in terms of the mean nearest-neighbor distance $d$ and the number of particles $N$ leads to
\begin{equation}
 	c(a,h) = C_{\text{local}}\left(\frac{a}{d}\right)^3 N+C_{\text{grids}}\left(\frac{a}{h}\right)^3 \left(\frac{d}{h}\right)^3 N.
 	\label{eq:cost2}
\end{equation}
Replacing $a$ using \autoref{eq:optimal-cutoff} and setting the derivative of \autoref{eq:cost2} to zero yields
\begin{equation}
	h = \left( {\frac{C_{\text{grids}}}{C_{\text{local}}} \frac{\left(n+14\right)}{n}} \right)^{1/6}d
	\label{eq:optimal-spacing}
\end{equation}
for the optimal spacing of the finest grid. Hence, the optimal spacing is
independent of the desired accuracy, while knowing the ratio
${C_{\text{grids}}}/{C_{\text{local}}}$ is important for selecting both $h$ and $a$, as they are coupled by \autoref{eq:optimal-cutoff}. One can estimate $C_{\text{local}}$ and $C_{\text{grids}}$ by curve fitting of the timings presented in \autoref{figure:short-long-range-timings} and \autoref{figure:mixing-timings}.

\subsection{Theoretical Complexity of the Multilevel Summation Method}
\label{section: theoretical complexity}
\autoref{eq:cost2} shows that the terms dominating the cost of the MLS method scale linear in $N$.
We have thus far not discussed the complexity of anterpolation and interpolation, as well as restrictions and prolongations.
Because of the restricted support of the nodal basis function, only nearby grid points are involved in the mappings in all those steps, yielding them linear in nature.
In addition, one can show for restrictions and prolongations that the total
cost does not exceed 8/7 times the cost of restriction from or prolongation to
the finest grid.
Thus, when the number of grids is large enough that the quadratically scaling cost of the last grid is negligible, the MLS algorithm should be linear in time.

\begin{figure}[ht]
  \begin{center}
    \includegraphics{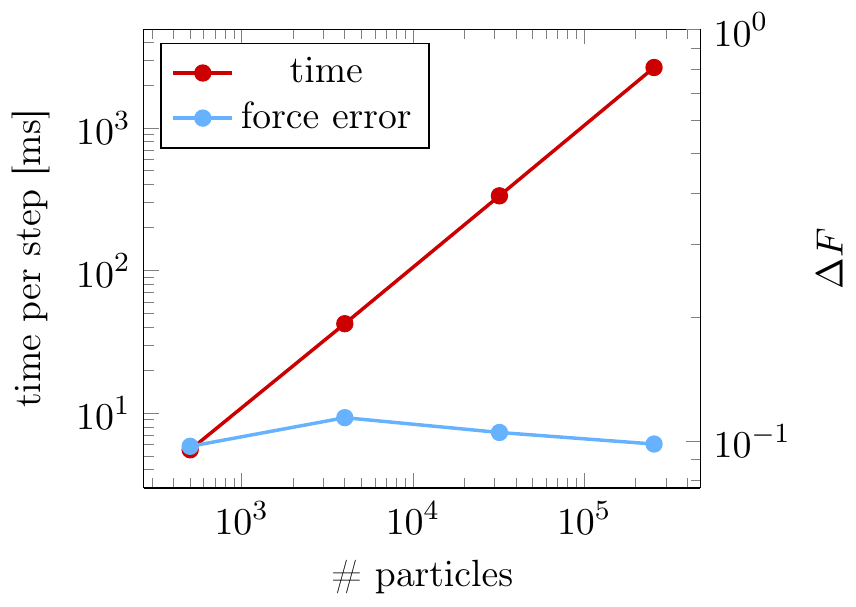}
    \caption{Total timings per step and error in forces for 500, 4000, 32000 and 256,000 particles.}
    \label{figure:linear-scaling}
  \end{center}
\end{figure}

To verify the theoretical analysis, we carried out simulations with 500,
4,000, 32,000 and 256,000 particles for the interfacial system discussed in \autoref{section: implementation}. The overall and slab densities, the cutoff $a$, the cell width of the linked cells, and the finest-grid spacing $h$ were all kept constant. The number of grids was increased by one from system to system to keep the all-to-all calculation cost on the last grid fixed.
The timings of these simulations are presented in \autoref{figure:linear-scaling}, that convincingly shows the linear scaling, while the error in the forces at the initial step is almost invariant for the different systems.

%%% Local Variables:
%%% mode: latex
%%% TeX-master: "DanielPaper01"
%%% End:

% Conclusions

\section{Conclusions}
\label{section: conclusions}

We have extended the multilevel summation method, originally applied to Coulombic interactions, to handle long-range dispersion potentials for molecular dynamics simulations of multiphase systems.
A prototype of the MLS algorithm for dispersion has demonstrated the linear scaling in the number of particles.
For the given choices of the splitting function $\gamma$ used here, currently a smoothing function with $C^2$ continuity and an interpolation order of three achieve the best results.
We have derived conservative bounds for the error, and estimates for the performance of the MLS method.
The error is shown to depend strongly on the cutoff $a$ as well as the spacing $h$ of the finest grid, with a lesser dependence on the number of grids.
The parameters $a$ and $h$ dominate the performance estimates too; consequently,
we have also derived qualitative estimates for the optimal choice of both $a$ and $h$ for given systems.
With improvements in the error bounds as well as algorithmic performance, quantitative predictions can be offered.

We further note that this work represents the next stage in the evolution of
long-range dispersion methods, beginning with traditional Ewald
methods\cite{Veld:2007ip} and continuing with Ewald mesh methods.\cite{Isele-Holder2012}
As part of this process, the complexity of the calculation has been reduced from $\mathcal{O}(N^2)$ to $\mathcal{O}(N \log N)$ and now finally to $\mathcal{O}(N)$.
Choosing between these methods, however, will require further study along the
lines of the previous work of Pollock and Glosli\cite{Pollock:1996uz} with
respect to efficiency, and that of Deserno and
Holm\cite{Deserno:1998ug,Deserno:1998wb} with respect to accuracy.
Accurately describing the behavior of molecular systems across interfaces is of such broad importance in modern research and commercial applications that we believe that such effort is of more than esoteric interest, and can instead lead to significant new capabilities within the molecular modeling community.
\footnote{The prototype code is available for download as supporting information (\url{https://github.com/daniel-tameling/MLS_Disp_Prototype}); a production version of this code will be released as part of the LAMMPS molecular dynamics package.}

%%% Local Variables:
%%% mode: latex
%%% TeX-master: "DanielPaper01"
%%% End:

\section*{Acknowledgments}

The authors would like to thank Steve Plimpton, Paul Crozier, and Stan Moore for helpful discussions and information regarding the LAMMPS implementation of MLS for Coulombic potentials.
Financial support from the Deutsche Forschungsgemeinschaft (German Research Foundation) through Grant GSC 111 is gratefully acknowledged.

\bibliographystyle{aipnum4-1}

\end{document}